\documentclass[aps,pra,floatfix,footinbib,preprintnumbers,superscriptaddress,reprint,longbibliography
]{revtex4-2}

\usepackage{amsfonts}       % blackboard math symbols
\usepackage{nicefrac}       % compact symbols for 1/2, etc.
\usepackage{microtype}      % microtypography
\usepackage{ulem}           % for `sout`
\usepackage{lipsum}
\usepackage{graphicx}
\usepackage{amssymb}
\usepackage[dvipsnames]{xcolor}
\usepackage{mathtools}
\usepackage{natbib}
\AtBeginDocument{\hypersetup{linkcolor=Blue,urlcolor=Blue,citecolor=Blue}}
\usepackage[colorlinks=true,hypertexnames=false]{hyperref}
\usepackage[capitalise]{cleveref}
\Crefname{section}{Sec.}{Secs.}
\usepackage[norelsize, linesnumbered, ruled, lined, boxed, commentsnumbered]{algorithm2e}

\usepackage{amsthm}
\newtheorem{theorem}{Theorem}[section]
\newtheorem{corollary}{Corollary}[theorem]

\newtheorem{proposition}[theorem]{Proposition}

\newcommand{\Tr}{\text{Tr}}

\newcommand{\edit}[1]{#1}
\newcommand{\editandy}[1]{#1}

\newcommand{\be}{\begin{equation}}
\newcommand{\ee}{\end{equation}}
\newcommand{\nn}{\nonumber}
\newcommand{\lf}{\left}
\newcommand{\rt}{\right}
\newcommand{\norm}[1]{\left\lVert#1\right\rVert}

\newcommand{\inv}{^{-1}}

\newcommand{\sket}[1]{| #1 \rangle}

\newcommand{\pexp}{p'}
\newcommand{\pideal}{p}
\newcommand{\papp}{\tilde{p}}
\newcommand{\Ro}{\textsf{R}}
\newcommand{\Rof}[1]{\textsf{R}_{#1}}
\newcommand{\RoTf}[3]{\textsf{R}_{#1}^{(#2,#3)}}
\newcommand{\er}{q}
\newcommand{\Rd}{\Rof{0}}
\newcommand{\So}{\textsf{S}}
\newcommand{\Po}{\textsf{P}}

\newcommand{\erfull}{\epsilon}

\begin{document}
\title{Perturbative readout error mitigation for near term quantum computers}
\author{Evan~Peters}
\email{e6peters@uwaterloo.ca}
\affiliation{Fermi National Accelerator Laboratory, Batavia, IL 60510}
\affiliation{Institute for Quantum Computing, University of Waterloo, Waterloo, Ontario, N2L 3G1, Canada}
\affiliation{Department of Physics, University of Waterloo, Waterloo, Ontario, N2L 3G1, Canada}
\author{Andy C.~Y.~Li}
\affiliation{Fermi National Accelerator Laboratory, Batavia, IL 60510}
\author{Gabriel~N.~Perdue}
\affiliation{Fermi National Accelerator Laboratory, Batavia, IL 60510}

\preprint{FERMILAB-PUB-21-233-QIS}

% \keywords{quantum computing}

\date{\today}

\begin{abstract}
Readout errors on near-term quantum computers can introduce significant error to the empirical probability distribution sampled from the output of a quantum circuit. These errors can be mitigated by classical postprocessing given the access of an experimental response matrix that describes the error associated with measurement of each computational basis state.
However, the resources required to characterize a complete response matrix and to compute the corrected probability distribution scale exponentially in the number of qubits $n$.
In this work, we modify standard matrix inversion techniques using perturbative approximations with significantly reduced complexity and bounded error when the likelihood of high order bitflip events is strongly suppressed. Given a characteristic error rate $\er$, \edit{we discuss a method to recover} the probability of the all-zeros bitstring $\pideal_0$ by sampling only a small subspace of the response matrix before inverting readout error resulting in a relative speedup of $\text{poly}\left(2^{n} / \big(\begin{smallmatrix} n \\ w \end{smallmatrix}\big)\right)$, which we motivate using a simplified error model for which the approximation incurs only $O(\er^w)$ error for some integer $w$. 
We then provide a generalized technique to efficiently recover full output distributions with $O(\er^w)$ error in the perturbative limit. These approximate techniques for readout error correction may greatly accelerate near term quantum computing applications. 
% We also investigate the further enhanced speed-ups in the special case of recovering the the probability of the all-zeros bitstring, which is a common target observable for several near-term quantum algorithms.
\end{abstract}

\maketitle

% % % % % % % % % % % % % % % % % % % % % % % % % % % % % % % % % % % % % % % % 
\section{Introduction}

While quantum computing will potentially provide an exponential speed up in solving certain problems, noisy intermediate-scale quantum (NISQ) \cite{Preskill2018quantumcomputingin} devices are subject to high error rates that must be mitigated in order to extract useful information from the quantum processors. Readout error is unique among the standard sources of decoherence since it is well modelled by a \textit{classical} stochastic process and is therefore entirely reversible by classical post-processing. In the simplest approach, the effects of readout error can be reversed by inverting \edit{a} \textit{response matrix} $\Ro$ that relates pre-measurement computational basis states to bitstrings sampled by the measurement, provided that $\Ro$ is nonsingular and accurately characterizes the readout error dynamics.

Previous works in readout error mitigation typically involve some variation of inverting a Markovian process \cite{Neeley_2010,Willsch_2018,Dewes_2012,Chen_2019,Gong_2019,havlicek_2019,Wei_2020}. The goal of these post-processing techniques is to recover the entire probability mass function $p(x)$ over bitstrings $x\in\{0, 1\}^n$. However, these techniques are generally not scalable as they require experimental characterization of a response matrix followed by an (approximate) matrix inversion step, requiring device time and computing resources that grow exponentially in $n$. Bayesian Iterative Unfolding \cite{Urbanek_2020,nachman_unfolding_2020} avoids the latter hurdle by approximating the matrix inversion and readout rebalancing \cite{hicks_readout_2021} improves on the accuracy of readout error correction for recovering high-weight bitstrings by biasing measurements based on some prior knowledge of the support of $p(x)$ on $\Ro^{2^n}$. However, unless error mitigation is applied to recover a specific observable \cite{funcke2020measurement,bravyi2020mitigating,PhysRevA.105.032620} both techniques still generally require an exponentially large device time to characterize the response matrix. Furthermore, with limited exceptions (e.g., ref.~\cite{peters2021machine}), few of these techniques have been specialized for the case where only a single bitstring probability is desired, which requires significantly fewer resources to mitigate readout error.

In this work, we present \edit{a perturbative technique} for approximately correcting readout error on near-term quantum computers. \edit{Intuitively, the technique relies} on an assumption that the likelihood of a readout error event involving many simultaneous bitflips (for instance, observing the bitstring $1111$ after a computational basis measurement of the state $|0000\rangle$) is strongly suppressed in the number of simultaneous bitflips. This includes scenarios for which the bitflips are weakly correlated between different qubits and the individual bitflip rates are small, which is often the case on existing devices \cite{nachman_2021}. 

\edit{We introduce our technique by considering the task of recovering} the probability of the all-zeros bitstring $\pideal_0$ \edit{and show that this may be accomplished} using only a small submatrix of $\Ro$, and we provide numerical and theoretical evidence justifying this approximation. \edit{By tailoring experimental determination of $\Ro$ towards recovering a specific bitstring even in the presence of correlated readout errors, this approach offers a potential performance advantage over existing techniques designed to recover full distributions.}

We then present the general technique to approximately recover the full empirical bitstring probability distribution by perturbatively expanding $\Ro$ in terms of a characteristic readout error rate $\er$. \edit{This technique represents a middle ground between full matrix inversion of $\Ro$ and sparsity-based techniques. It is well-suited for mitigating readout error when both the strength of correlations between readout errors is known and the distribution $p(x)$ has non-trivial support on a large number of bitstrings (e.g. superpolynomial in $n$), such that the probability of observing each bitstring is influenced by the underlying probabilities for many other bitstrings that are close in Hamming distance.} Both \edit{variants of our technique} allow for probabilities to be \textit{approximately} corrected with the benefit of greatly reduced error correction overhead, and are therefore especially well suited for experiments in which readout error is not the limiting factor in the accuracy of the sampled probabilities.

% \begin{figure*}
%     \centering
%     \includegraphics[width=0.8\linewidth]{figures/process.pdf}
%     \caption{Visualization of the truncated readout error correction method of Equation~\ref{eq:p0_prime}. \textbf{a} Typical readout error correction involves sampling $\Ro$, computing $\Ro^{-1}$, and then correcting for the prior bitstring distribution $p$ given the observed $\pexp$. However if only a single bitstring probability is desired, the majority of $\Ro^{-1}$ is discarded. \textbf{b} The scheme we study in this work involves computing a truncated $\Ro_T$ for which $\Ro_T^{-1}$ is \textit{similar to} the truncated $\Ro^{-1}$ on the subspace used to recover $p(0)$. }
%     \label{fig:p0_recover}
% \end{figure*}

% % % % % % % % % % % % % % % % % % % % % % % % % % % % % % % % % % % % % % % % 
\section{Inverting readout error}

Given an $n$-qubit state represented by its density matrix $\rho$, the error in a projective measurement $\{|i\rangle \langle i|\}$ for $i=0, \dots, 2^n - 1$ over the computational basis states  can be modelled as a classical Markovian process \cite{geller_rigorous_2020}, which is described by \edit{the equation}
\begin{equation}\label{eq:bf_R}
    \pexp = \Ro \, \pideal.
\end{equation}
Here, $\Ro$ is a $2^n \times 2^n$ matrix \edit{with nonnegative entries} whose columns sum to one \edit{(i.e. a left stochastic matrix)}, $\pideal= \text{diag}(\rho)$ is a length-$2^n$ normalized array of probabilities measured in computational basis without measurement noise, and $\pexp$ is the length-$2^n$ array of observed (erroneous) bitstring probabilities. The response matrix $\Ro$ may be defined elementwise in terms of transition likelihoods,
\begin{equation}\label{eq:bf_prob}
\Ro_{ij} \equiv p(i|j) = p(i_1\dots i_n | j_1 \dots j_n)
\end{equation}
where $i,j\in\{0,1\}^n$ are length-$n$ bitstrings, and the notation $i_k$ is understood to refer to the $k$-th bit of $i$. If we are provided with an invertible $\Ro$, a basic prescription for correcting readout error is to compute 
\begin{equation}\label{eq:vanilla_inv}
\pideal = \Ro^{-1} \pexp
\end{equation}
In practice, $\Ro$ may be singular and a least squares approximation to the linear equation \eqref{eq:bf_R} may be used.

Even when $\Ro$ is invertible, computing  \cref{eq:vanilla_inv} in a general setting requires two distinct, resource intensive steps: (i) measuring the complete response matrix of bitstring transition probabilities using a diagnostic experiment to determine $\Ro$, with time complexity $\mathcal{O}(2^{n})$ and (ii) performing matrix inversion on $\Ro$, which can be as costly as $\mathcal{O}(2^{3n})$ \footnote{For simplicity, we will assume both inversion and multiplication of  generic $M\times M$ matrices inversion proceeds with complexity $\mathcal{O}(M^3)$. Optimized algorithms like Strassens' algorithm reduce this complexity but these considerations will not affect the \textit{relative} speedups that we present in this work.} \edit{Notably, recent works have explored more efficient sparsity-based techniques for mitigating readout error, for example inverting the response matrix in the subspace spanned by nonzero components of $p'$ \cite{PRXQuantum.2.040326,PhysRevA.106.012423}.}

A small infidelity in the readout error mitigation can usually be tolerated as a trade-off for an improved complexity scaling in many cases, for example, when the readout error is less significant compared to other sources of error such as decoherence. We now introduce heuristic techniques for reducing the resource requirements of both of these steps while incurring some small, controllable error that may be inferred under some mild assumptions about the structure of $\Ro$.

% % % % % % % % % % % % % % % % % % % % % % % % % % % % % % % % % % % % % % % % 
\section{Recovering the probability all-zeros bitstring}
\label{sebsec:all-zeros_bitstring}

We first study a \edit{special case of our technique} for complexity reduction \edit{when one only desires} to determine the probability of the all-zeros bitstring $\pideal_0 = \Tr(|0^n\rangle \langle 0^n|\rho)$. This scenario is relevant for near term algorithms such as quantum kernel methods \cite{Schuld2019a,havlicek_2019}, \edit{dual-state purification \cite{huo_2022}, qubit assignment on hardware \cite{peters_2022}, and quantum  circuit learning \cite{Mitarai2018,Khatri_2019}}. In this context, \cref{eq:vanilla_inv} can be cast in the form of a dot product,
\begin{equation}\label{eq:p0_simple}
    \pideal_0 = r \cdot \pexp
\end{equation}
where the vector $r \in \mathbb{R}^{2^n}$ is defined elementwise as $r_i = (\Ro^{-1})_{0i}$. One would expect that a simplified readout error mitigation can be performed to recover the observable $p_0$ with both tight error bounds and greater efficiency than for recovering the full distribution $p$, since only the subspace of $\Ro^{-1}$ that describes likely transitions into and out of $0^n$ is relevant. Assuming that the probability of a bitstring transition falls monotonically in the number of indiviual bits flipped, this subspace corresponds to the set of probabilities $(\Ro)_{ij}$ for which $i$ and $j$ are low-weight strings.

% That is, rewriting \cref{eq:bf_R} in terms of the relevant components for readout error on the all-zeros bitstring as
% \begin{equation}
%     p_0' = p_0 - \underbrace{\sum_{x \neq 0}R_{x0} p_0}_\text{outflux $\leftarrow 0$} + \underbrace{\sum_{x \neq 0} R_{0x} p_x}_\text{influx $\rightarrow 0$}
% \end{equation}

Following this intuition, \edit{our proposed technique works by correcting} $\pideal_0$ using the inverse of a projection of $\Ro$ onto the subspace of low-weight basis vectors. The weight of a binary bitstring $x=x_1 x_2 \dots x_n \in \{0,1\}^n$ is defined as
\begin{equation}\label{eq:weight}
    w(x)= \sum_{i=1}^n x_i
\end{equation}
which is the number of 1's appearing in $x$. We denote the set of all bitstrings with weight less than $w$ as 
\begin{equation}
    S_w = \{ x:  x\in \{0,1\}^n, \, w(x) \leq w\} \subseteq \{0,1\}^n.
\end{equation}

We then define the weight projection operator $P_w: \mathbb{R}^d \rightarrow \mathbb{R}^{|S_w|}$ that projects vectors onto the subspace spanned by basis vectors whose binary index is in $S_w$. This is equivalent to the action
\begin{equation}\label{eq:proj_w}
P_{w} \hat{e}_j =
\begin{cases}
	\hat{e}_j,& \text{if } j \in S_{w}, \\
	0,              &  \text{otherwise}.
\end{cases}
\end{equation}
where $\hat{e}_j$ is the $j$-th unit vector. Then, a $d \times d$ matrix $A$ may be projected onto the same subspace by the operation \edit{$P_w A P_w$}. Defining \edit{$\Ro_T =  P_w \Ro P_w$} and the first row of its inverse as $(r_T)_i = (\Ro_T^{-1})_{0i}$ by analogy with \cref{eq:p0_simple}, our goal is to demonstrate that for some choice of $w < n$ and mild assumptions about the structure of $\Ro$, we can compute 
\begin{equation}\label{eq:p0_prime}
    \papp_0 = r_T \cdot \pideal_T'
\end{equation}
as a close approximation to $\pideal_0$, where $\pexp_T = P_w \pexp$. \Cref{eq:p0_prime} simply uses the first row $r_T$ of the inverse of a truncated response matrix in place of $r$ applied to a truncated observed probability vector $\pexp_T$. Applying \cref{eq:p0_prime} consumes a significantly smaller response matrix matrix $\Ro_T$ which may be understood conceptually as the top-left submatrix of $\Ro$ with rows and columns resorted by index weight. $\Ro_T$ has dimensions $t_w \times t_w$ given by the sum over binomial coefficients 
\begin{equation}
    t_w = \sum_{j=0}^w \begin{pmatrix} n \\ j \end{pmatrix}
\end{equation}
and so the readout error correction for $\tilde{p}_0$ may be carried out by sampling $\Ro_T$ from a quantum processor with complexity $\mathcal{O}(t_w)$ and then computing $\Ro_T^{-1}$ with complexity $\mathcal{O}(t_w^3)$. In this work, we will neglect the effects sampling error in both $R$ and $p'$, which introduces a constant overhead for readout error correction as a function of the strength of noise on the device \cite{bravyi2020mitigating}. In the absence of statistical effects \edit{(such that Eq.~\ref{eq:p0_simple} is satisfied) the error of our method is  given as
\begin{align}
    | r_T \cdot \pexp_T - p_0| &= | r_T \cdot \pexp_T - r \cdot \pexp|
\end{align}
}

This approach therefore introduces error from two different sources: the first kind of error results from computing $\pideal$ after discarding bitflip events involving more than $w$ simultaneous relaxations and excitations, while the second kind of error results is due to the truncation approximation \edit{$\Ro_T^{-1}$ - $P_w \Ro^{-1} P_w$}. To motivate our technique, we proceed study situations for which this difference vanishes and provide the resulting bounds on $| r_T \cdot \pexp_T - r \cdot \pexp|$ for progressively looser restrictions on the structure of $\Ro$. 

% % % % % % % % % % % % % % % % % % % % % % % % % % % % % % % % % % % % % % % % 
\subsection{Exact bounds for a relaxation-only model}
It is convenient to use the convention that vectors and matrices be sorted according to the weight of the binary representation of the index, with indices of equal weight sorted arbitrarily. For example with $n=3$, this has the effect of rearranging the vector of readout probabilities such that
\begin{equation}
    \pideal = (p_{000}, p_{001}, p_{010}, p_{100}, p_{110}, \cdots )^{T}
\end{equation}

We now consider an instructive toy model for readout error for which an analytical upper bound on the error $|\pideal_0 - r \cdot \pexp|$ may be derived exactly. In this model, $\Ro$ is both overly simplified and trivially invertible, but the analysis will provide insight into approximations for situations where $\Ro$ has a more complex structure. The model for readout error that we \edit{study analytically} is an extreme example of asymmetric readout error described by a response matrix of the form
\begin{align}\label{eq:type1}
    \Ro &= \bigotimes_{k=1}^n Q_k\\
        Q_k&= \begin{pmatrix} 
            1 & \er \\
            0 & (1-\er)
        \end{pmatrix} 
\end{align}
\edit{for $0\leq \er < 0.5$}. We can make very strong arguments about readout error arising from this model. 

\begin{proposition}\label{lemma:1}
Let $\Ro$ be defined as in \cref{eq:type1}. For any fixed projector $P_{w}$ satisfying $P_{w}^2=P_{w}$, define \edit{$\Ro_T =P_{w} \Ro P_{w} $}. Then,
\begin{equation}
   P_{w} \left(\Ro^{-1}\right) P_{w} =   \left(\Ro_T\right)^{-1} 
\end{equation}
\end{proposition}
In other words, for this definition of $\Ro$ the inverse of the projected response matrix $\Ro_T$ is equal to a projection of  $\Ro^{-1}$. This is a straightforward property of the kinds of upper triangular matrices we are interested, but an intuitive proof is provided in Appendix~\ref{app:a}. The following theorem applies \edit{Proposition~\ref{lemma:1}} to show that we can compute only a very small subspace of $\Ro$ and invert that subspace to apply readout error correction to recover $\pideal_0$ with error that is exponentially suppressed in our choice of truncation weight $w$.

\begin{theorem}\label{thm:1}
Let $\Ro$ be defined as in \cref{eq:type1}. Then the error introduced by correcting readout error using a truncated response matrix is bounded by:
\begin{equation}
   |  r_T \cdot p_T' -  r \cdot p'| \leq \left(2q\right)^{w+1}
\end{equation}
where $r_T$ and $r$ are defined elementwise as $(r_T)_i = (\Ro_T^{-1})_{0i}$ and $r_i = (\Ro^{-1})_{0i}$.
\end{theorem}
% \leq \mathcal{O}(q^w)
The proof is given in Appendix~\ref{app:b}. We remark that this is the tightest possible bound given the structure assumed of $\Ro$ that does not incorporate additional information about the readout probability distribution over the truncated subspace. Theorem~\ref{thm:1} makes a simple but powerful observation that given the restricted noise model we have considered, one can apply readout error using the inverse of a projected matrix $\Ro_T$ such that the truncation error introduced is exponentially suppressed in $w$. This bound becomes quite weak in the limit that $\er\rightarrow 0.5$ since the dimension of the truncated matrix itself grows combinatorially in $w$.  Conversely, for $\er \ll 1$ such that $\er^2 \rightarrow 0$, this result guarantees that a matrix projected onto the $w=1$ weight subspace with size \textit{linear in $n$} can recover the probability of $0$ with an accuracy almost as good as using the exponentially large $\Ro$. Since this result is based only on the structure of $\Ro$ and not the value of elements contained therein, we can immediately lift some of the restrictions on constructing $\Ro$.

\begin{corollary}\label{corollary:2}
Let $\Ro$ have a tensor structure composed of distinct individual qubit response matrices of the following form: 
\begin{equation}
    \Ro = \bigotimes_{k=1}^n \begin{pmatrix} 
            1 & \er_k \\
            0 & (1-\er_k)
        \end{pmatrix}
\end{equation}
for $0\leq \er_k < 0.5$. Then 
\begin{equation}
   |  r_T \cdot p_T' -  r \cdot p'| \leq \left(2 q_{max}\right)^{w+1}
\end{equation}
where $\er_{max} = \max_k \{\er_k\}$.

\end{corollary}
This is shown in Appendix~\ref{app:b}. Corollary~\ref{corollary:2} expands on the intuition of Theorem~\ref{thm:1}: If $k$-th order simultaneous bitflip events are suppressed exponentially in $k$, then we need only sample a submatrix of $\Ro$ to perform good readout correction. We can further extend this line of reasoning to its practical limit in a somewhat less rigorous way. Suppose $\Ro$ is \textit{any} response matrix that allows only for ``relaxation'' events, that is $\Ro$ may be defined elementwise as 
\begin{equation}
    \Ro_{ij} = 
    \begin{cases}
         p(i|j) & \text{for } w(i) < w(j) \text{ or } i=j,\\
        0 & \text{otherwise. } \\
     \end{cases}
\end{equation}

If we assume that the probability of a relaxation event is suppressed exponentially in the number of simultaneous bitflips, i.e. $p(i|j) \leq \mathcal{O}\left(\er^{w(j) - w(i)}\right)$ for some characteristic rate $\er$ then the above bounds still hold in the approximate sense:
\begin{equation}\label{eq:loose}
   |  r_T \cdot p_T' - r \cdot p'| \lesssim \mathcal{O}\left( 2\er^{w + 1} \right)
\end{equation}
This follows directly from Theorem~\ref{thm:1}; each entry with magnitude exactly $(1-\er)^{w(x)} \er^{w(y)}$ in the strictly upper triangular part of $\Ro$ can be replaced with approximate term with order $\mathcal{O}\left((1-\er)^{w(x)} \er^{w(y)}\right)$. As this modification does not affect the structure of $\Ro$, similar nilpotency and series expansion arguments that lead to Theorem~\ref{thm:1} may be applied by substituting $\er \rightarrow \mathcal{O}(\er)$. In this situation $\Ro$ can no longer be decomposed and therefore $\Ro^{-1}$ can no longer be efficiently computed as $\bigotimes_k Q_k^{-1}$ using individual qubit response matrices $\{Q_k\}_{k=1}^n$. This extension also marks a departure from Corollary~\ref{corollary:2} by relaxing the assumption that $\Ro$ has a tensor structure, and therefore accommodates weakly correlated readout errors. Despite this structural change, the projected $\Ro_T^{-1}$ constructed from events with order no greater than $w$ still serves as a useful surrogate for $\Ro^{-1}$ if only elements in the first row of $\Ro\inv$ are desired, and provides some justification for extending the reasoning of \cref{eq:p0_prime} to the more general case.

% We have introduced a technique for readout error correction for recovering specifically $\pideal_0$ and validated this technique for a limited but tractable readout error model. The reasoning was based on a plausible assumption that there exists some characteristic single-qubit readout error rate of $\er$ and that events involving many simultaneous bitflips can be safely ignored in studying readout error dynamics. 

% % % % % % % % % % % % % % % % % % % % % % % % % % % % % % % % % % % % % % % % 
\section{Perturbative mitigation for recovering the full distribution}
\label{sec:pert_full}

In the previous section, projecting $\Ro$ onto a subspace of low-weight indices was motivated by a model for readout error that penalizes the transition of high-weight bitstrings into $0^n$. This reasoning can be generalized to recovering the full bitstring distribution $\pideal$ more efficiently, assuming an error model that penalizes transitions between any two bitstrings that differ by a large hamming weight. This is a practical model even even when there is correlated readout error between different qubits, provided the correlation strength is not comparable to the characteristic rate. This model is further motivated by the observation that correlations in readout error are likely to be strong only among qubits that are physically adjacent on a device, for example nearest neighbors on a two-dimensional grid of superconducting qubits \cite{Geller_2021}.

To proceed, we assume there is a characteristic rate $\er$ that describes the probability of any given single bitflip event. For each $j=1, \dots, n$, we define  a sparse $2^n \times 2^n$ off-diagonal matrix $\Rof{j}$ whose entries are of magnitude $\mathcal{O}(1)$. Then, without loss of generality, we define $\Ro$ with respect to a series structure such that
\be\label{eq:R_decompose}
\Ro = \Rd + \sum_{j=1}^{2^n} \er^{j} \Rof{j},
\ee
where each $\Rof{j}$ contains a subset of elements of $\Ro$ according to some pairwise comparison function $s$,
\begin{equation}\label{eq:Rj_nm}
    (\Rof{j})_{nm} = \begin{cases}
        (\Ro)_{nm} & \text{if }s(n, m) = j, \\
        0 & \text{otherwise.}
    \end{cases}
\end{equation}

In this work, we will focus on the specific choice
\begin{equation}
    s(n, m) = w(n \oplus m)
\end{equation}
where $w$ is the weight function of \cref{eq:weight} and $\oplus$ denotes the bitwise modulo-2 sum. Then, $\Rd$ consists of the diagonal matrix elements of $\Ro$ of all orders of $\er$ and $\Rof{j}$ consists of the off-diagonal terms of the order $q^j$ describing all likelihoods involving bitflips whose weights differ by $j$. Note that under this definition, $\Rd$ will describe all bitflip events of even order in which the observed bitstring is identical to the prior bitstring, and similarly for $j=1, \dots, n$ so that $\Rof{j}$ does not necessarily characterize events involving exactly $j$ bitflips. Even if $\Ro$ is not well modelled by this choice of decomposition, such as in cases involving strongly correlated readout errors, it still may be possible to define $\Rof{j}$ to include all events with probability on order $q^j$. This will require significant prior knowledge about the scale of readout errors on the device and is out of scope for this work. 

\begin{figure}
    \centering
    \includegraphics[width=\linewidth ]{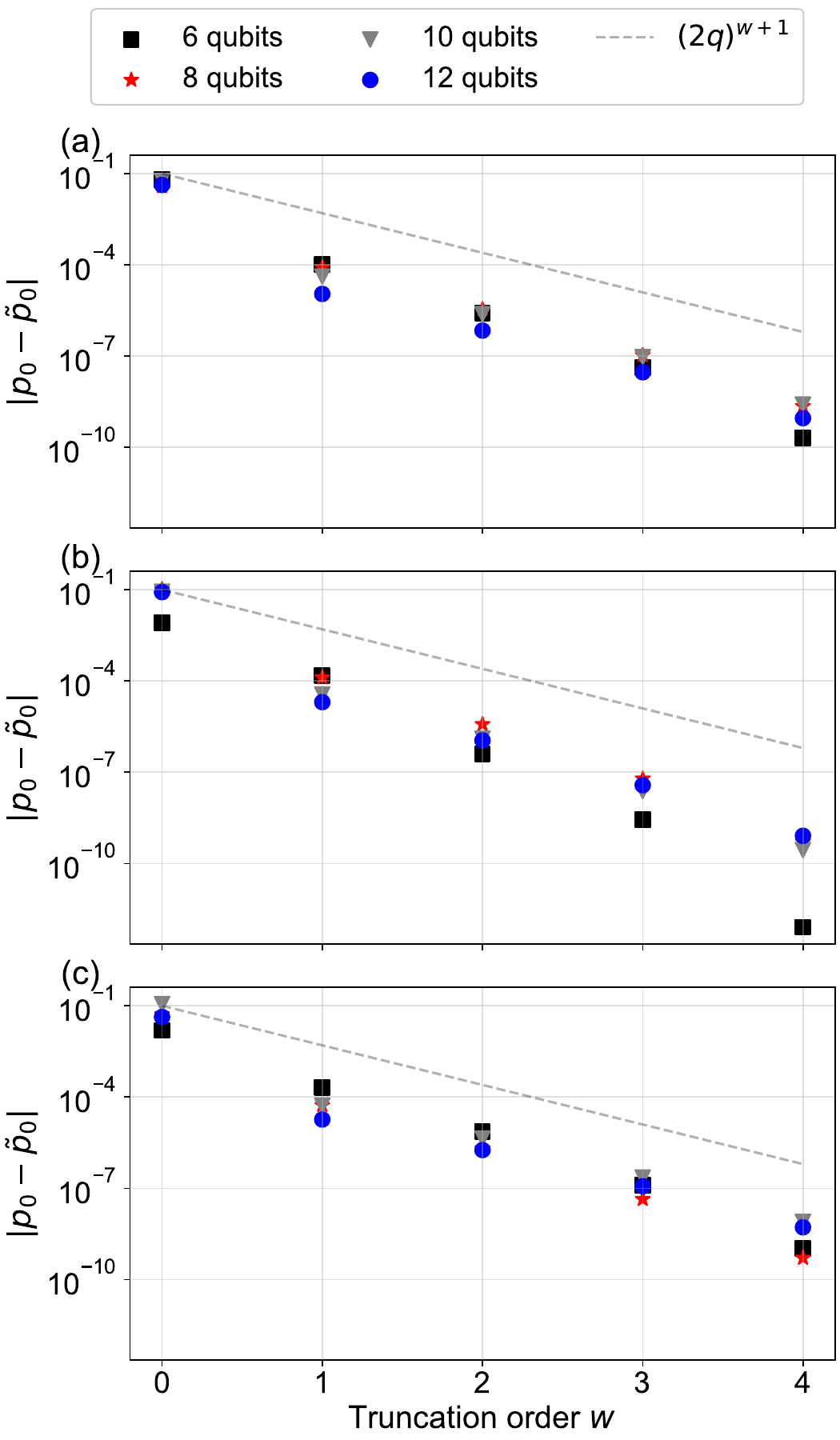}
    \caption{Performance of all-zeros readout error mitigation given by \cref{eq:p0_prime} compared for different prior distributions. (a) The Gaussian prior is centered at $0$ with $n$-bit overflow for all bitstrings with value less than $2^{n-1}$, i.e. $p_j \propto \exp\left( (x_j - 0.5)^2 / \sigma^2 \right)$ where $x_j \equiv 2^{-n} \left((j + 2^{n-1}) \text{ mod } 2^n\right)$ and $\sigma=0.25$. This distribution is adversarial to recovering $p_0$ as it has significant support on the high-weight subspace. (b) The truncated Gaussian is given by the same distribution without overflow ($x_j = j \cdot 2^{-n}, \sigma=0.25)$ and renormalized, and the (c) uniform distribution is $p_j=2^{-n}$. In all plots, $w=0$ is defined to correspond to the uncorrected probability $\pexp_0$. The dashed line indicates the bound of \cref{eq:loose} derived for a relaxation-only model, which we observed was not violated even for the more general error model of \cref{eq:general_R}. }
    \label{fig:numeric1}
\end{figure}

The form of \cref{eq:R_decompose} suggests that the $\Rof{j}$ corresponding to small $j$ will dominate the effects of dynamics error. Applying this intuition, we expand the inverse of \cref{eq:R_decompose} as a series,
\begin{align}\label{eq:R_series}
\Ro\inv & = \Rd\inv + \sum_{k=1}^{\infty} \lf(- \sum_{j=1}^{2^n} \er^{j} \Rd\inv \Rof{j} \rt)^k \Rd\inv
.
\end{align}
Truncating both series to the order of $\er^{w}$ results in 
\be
\label{eq:R_inverse_app}
\Ro\inv = \lf[ 1  + \sum_{k=1}^{w} \lf(- \sum_{j=1}^{w} \er^{j} \Rd\inv \Rof{j} \rt)^k \rt] \Rd\inv
+ \mathcal{O} (\er^{w + 1})
.
\ee
\edit{Note that this expression contains some terms of order higher than $\er^{w}$ but the error introduced by the truncation remains bounded by $\mathcal{O} (\er^{w + 1})$.}
Applying this expression to $\pexp$, we arrive at an approximate probability distribution $\papp$ given by
\be
\label{eq:app_prob_full}
\papp = \lf[ 1  + \sum_{k=1}^{w} \lf(- \sum_{j=1}^{w}  \Rd\inv \mathcal{R}_{j} \rt)^k \rt] \Rd\inv \pexp
.
\ee
where we have introduced $\mathcal{R}_j \equiv q^j \Rof{j}$ as the matrix of elements describing order-$j$ transitions sampled directly from an experimental response matrix (for which knowledge of the rate $\er$ is not strictly necessary).  This result can be viewed as a generalization of the \edit{specialized task described} in ~\cref{eq:p0_prime}, which we discuss in Appendix~\ref{app:c}.

The implementation of the perturbative readout error mitigation to recover an empirical distribution over bitstrings sampled from a quantum computer subject to measurement error is described by the following pseudocode.

\IncMargin{1em}
\begin{algorithm}[H]
	\SetAlgoLined
	\LinesNumbered
	\SetKwInOut{Input}{Input}
	\Input{$\pexp, \{\mathcal{R}_j\}_{j=0, 1, \cdots, w}$}
	
	$\So \leftarrow - \sum_{j=1}^{w} \Rd\inv \mathcal{R}_j$ \;
	$v \leftarrow \Rd\inv \pexp$ \;
	$\papp \leftarrow v$ \;
	\For{$k \leftarrow 1$ \KwTo $w$}{
		$v \leftarrow  \So v$ \;
		$\papp \leftarrow  \papp + v$ \;
	}
	\caption{Perturbative mitigation for the full distribution}
	\label{alg:perturbative_mit_full}
\end{algorithm}
\DecMargin{1em}

\edit{
We note that ref. \cite{wang2021measurement} also employed series approximations for computing $\Ro\inv$ but the implementation is otherwise unrelated to the technique described here.
}

If $\Ro^{-1}$ exists, then the Neumann series introduced in \cref{eq:R_series} converges only if $\norm{\sum_{j=1}^w \Rd^{-1} \mathcal{R}_j} < 1$ which determines whether Algorithm~\ref{alg:perturbative_mit_full} can be applied in its given form. If this condition is met, then the error $\erfull = \norm{\pideal - \papp}_2$ introduced by Algorithm~\ref{alg:perturbative_mit_full} is concentrated in the $(w+1)$-th order terms, resulting in an approximate error given by $\erfull \lesssim 2 \er^{w+1} + \mathcal{O}( \er^{w+2} )$ where we have applied the slightly stronger assumption that $\norm{\Rd\inv\Rof{j}} \le 1$. To reach an accuracy with an error $\erfull$, we need to implement the algorithm with an order at least
\be
w \geq  \left\lceil\frac{  \ln \erfull\inv + \ln 2}{\ln  \er\inv  }\right\rceil - 1.
\ee

The complexity of this technique is dominated by matrix products involving $\So$ in Algorithm~\ref{alg:perturbative_mit_full}. If $\So$ is sparse with $s$ being roughly the fraction of elements that are non-zero, the algorithm requires $w$ matrix product operations resulting in approximate time complexity given by
\begin{equation}\label{eq:complexity}
    O(s_w w M^3 ).
\end{equation}
We now compute the sparsity of $\mathcal{R}_w$ to determine the relative speedup of this technique over standard matrix inversion. The nonzero elements of $\mathcal{R}_j$ occur at all index pairs $(x, y)$ satisfying $x \oplus y = j$, where $x,y\in B$ and $B= \{0,1\}^n$. The number of pairs $(x, y)$ that satisfy this condition is equivalent to the number of strings $x$ satisfying $x = z \oplus y$, where $z\in B_j$ and $B_j = \{s: s\in B, w(s)=j\}$ is the set of all weight-$j$ bitstrings. This number is given by $|B|\times |B_j|$, and so the sparsity factor $s_w$ describing the number of nonzero terms in $\mathcal{R}_w$  may be computed directly as $2^{-2n} \times |B|\times |B_j|$, or
\begin{equation}\label{eq:sw}
    s_w = 2^{-n} \begin{pmatrix}n \\ w\end{pmatrix}.
\end{equation}
\edit{This sparsity is computed using the union of Hamming balls around each weight-$k$ bitstring, i.e. our construction implicitly assumes readout errors that are only weakly correlated such that transitions between bitstrings with large Hamming distance are suppressed.}

In the context of \cref{eq:complexity}, $s_w$ has the effect of replacing one term proportional to $2^n$ with a term proportional to $w\big(\begin{smallmatrix} n \\ w \end{smallmatrix}\big)$. The core speed-up therefore comes generating $\tilde{p}$ in Algorithm~\ref{alg:perturbative_mit_full} by a series of sparse matrix-vector products using matrices with at most $s_w$ nonzero terms. If convergence of the Neumann series of \cref{eq:R_series} is not guaranteed, a perturbative technique may still be useful. In this case, an experimentalist would still measure the set $\{\mathcal{R}_j\}$ to a desired truncation point $w$, and then directly invert the resulting approximation to $\Ro$ to recover
\begin{equation}
    \papp = \left(\sum_{j=1}^w \mathcal{R}_j\right)\inv \pexp
\end{equation}

This may result in significant speedup compared to sampling the full $\Ro$, but incurs additional computational cost to compute a standard matrix inverse, and we explore this tradeoff in Appendix~\ref{app:d}.  This approach might therefore be compatible with other techniques that avoid directly computing a matrix inverse (for instance, Bayesian iterative unfolding \cite{Urbanek_2020,nachman_unfolding_2020,hicks_readout_2021}) or with  techniques for efficient inversion of sparse, banded matrices \cite{lipton1979generalized,KILIC2013126}.

\edit{
This approach allows us to safely ignore the small response matrix elements corresponding to higher-order terms. In experiments, each column of $\Ro$ may be estimated by preparing the state corresponding to that column and then computing the output bitstring distribution for a computational basis measurement, resulting in an estimate for the matrix elements of the corresponding column of $\Ro$. By discarding the higher-order terms, this distribution can be reliably determined with a number of shots like $n_{\mathrm{meas}} \sim 1/q^{2w}$. If a sufficient truncation order $w$ is known either from previous experiments or knowledge of the hardware design (e.g. Ref.~\cite{nachman_2021}), our technique saves resources by allowing the experimentalist to omit measurement of higher-order terms.
Moreover, the resource requirement may be reduced further if only a few bitstrings of the output distribution are required. This includes, for example, the all-zeros bitstring case discussed in \cref{sebsec:all-zeros_bitstring} and cases where only a specific qubit excitation sector is of interest due to symmetry. In these cases, we need only estimate columns of $\Ro$ by computing elements corresponding to the desired bitstring population (see Appendix~\ref{app:c}). This allows us to compute $\Ro$ using fewer experiments on quantum hardware.
}

% % % % % % % % % % % % % % % % % % % % % % % % % % % % % % % % % % % % % % % % 
\section{Numerical experiments}

We implemented the technique introduced in the preceding sections for a variety of prior distributions and readout error strengths. To avoid complications due to statistical uncertainty, we restrict ourselves to using $p$ and $\Ro$ that were simulated to floating point precision without introducing any sampling error. \edit{We generated each $\Ro$ randomly in the following manner: We constructed a tensor product of the form}
\begin{equation}\label{eq:general_R}
    \Ro = \bigotimes_{k=1}^n \begin{pmatrix} 1 - \eta_k & \epsilon_k \\ \eta_k & 1 - \epsilon_k\end{pmatrix}
\end{equation}
with $\epsilon_k, \eta_k \sim \text{Uniform}(0, \er)$, \edit{and then we randomly permuted each weight-$k$ subspace of $\Ro$ for $k=1,\dots,n-1$. The resulting matrix is not separable and therefore cannot be trivially inverted by inverting each term in the Kronecker product of~\cref{eq:general_R}. We applied the resulting linear map to a prior distribution $\pideal$ to generate $\pexp$.} \Cref{fig:numeric1} demonstrates the performance of \cref{eq:p0_prime} to correct for $p_0$ for different prior distributions. The error in the method is suppressed exponentially in the truncation order $w$, which is consistent with behavior that was analytically derived for a more restricted error model in \Cref{sebsec:all-zeros_bitstring}.
\edit{A similar exponential suppression of the error can be observed in experiments using response matrices measured on IBM QPUs (Appendix~\ref{app:numerics_QPU_results}). In Appendix~\ref{app:m3} we provide preliminary comparison of our method to the M3 technique of Ref.~\cite{PRXQuantum.2.040326}.}

\begin{figure}[!htp]
    \centering
    \includegraphics[width=\linewidth]{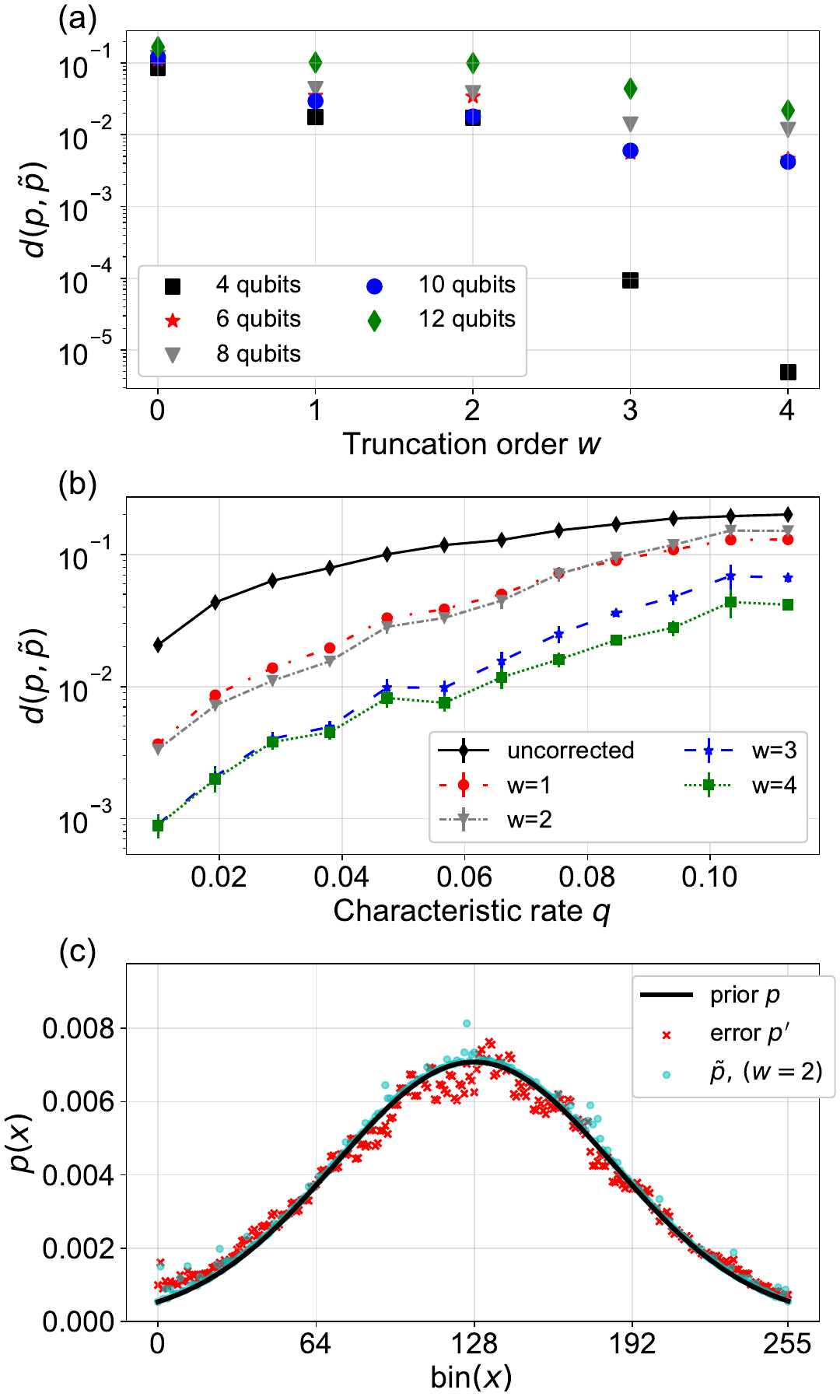}
    \caption{(a) Performance of Algorithm~\ref{alg:perturbative_mit_full} as a function of truncation order $w$, for fixed $\er=0.05$. The elements of the initial distribution $p$ were each drawn from $\text{Uniform}(0, 1)$ and then normalized. We define $w=0$ to represent the uncorrected case corresponding to $d(p, p')$. (b) The performance of the algorithm ($n=8$) diminishes with increasing $q$, which corresponds to the series approximation for $\Ro\inv$ diverging. We discuss the resulting limitations and workarounds for this behavior in Appendix~\ref{app:d}. (c) \edit{Visualization of applying correction to a Gaussian distribution for $n=8,w=2$ with a characteristic rate $\er=0.6$.}}
    \label{fig:numeric2}
\end{figure}

For Algorithm~\ref{alg:perturbative_mit_full}, we are interested in assessing the performance of the readout correction for recovering the full distribution $\pexp$ compared to $\pideal$. To compare the two distributions, we compute the trace distance (or L1 norm),
\begin{equation}\label{eq:trace_distance}
    d(p, q) = \sum_{j \in \{0,1\}^n} |p_j -q_j|.
\end{equation}
This has a useful interpretation in terms of computing expectation values of Hermitian operators. Let $O$ be an operator with $2^n$ real entries on the diagonal bounded by $O_{j}\in[-1,1]$. The expected value $\langle O \rangle = \Tr (O \rho)$ corresponding to the corrected distribution $\papp$ is estimated using the quantity
\begin{equation}
    E_O = \sum_{j\in\{0,1\}^n} O_{j} \papp_j.
\end{equation}
We can then show that $d(p, \papp)$ bounds the error incurred in $E_O$:
\begin{equation}
|\Tr (O \rho) - E_O| \leq d(p, \tilde{p}),
\end{equation}
and so $d$ serves as a natural comparison between output distributions that will be postprocessed to compute observables. \Cref{fig:numeric2} shows the performance of Algorithm~\ref{alg:perturbative_mit_full} for varying truncation orders, numbers $n$ of qubits and increasing noise rates $q$. For modest $\er$, the effects of readout error can be strongly suppressed using only a fraction of the resources required for inverting all of $\Ro$. As $n$ or $\er$ increases, the performance of the algorithm rapidly drops off as the series for truncated inverse requires significantly more terms to converge.
\Cref{fig:numeric2} also provides a visual example of  Algorithm~\ref{alg:perturbative_mit_full} applied using a second order truncation, \edit{which will generally consume} $\mathcal{O}(n^2)$ resources for sampling $\Rof{2}$ to recover $\tilde{p}$.

% % % % % % % % % % % % % % % % % % % % % % % % % % % % % % % % % % % % % % % % 
\section{Conclusion}

and may be appropriate  
which entails prohibitive experimental and computational overhead), (e.g. Ref.~\cite{PRXQuantum.2.040326}) that might 

well-suited 

We have proposed \edit{a technique} for approximately correcting readout error in quantum computers requiring significantly less overhead than traditional matrix inversion techniques, \edit{while still capturing enough of the readout error behavior to correct distributions with support on a large number of bitstrings that might present a challenge for sparsity-based techniques}. Such approximations are beneficial when the error in the correction scheme becomes negligible compared to other sources of device noise, and so \edit{this technique} may be useful for running quantum algorithms on near-term devices. We have justified \edit{the technique} in the perturbative regime and also provided numerical evidence suggesting that \edit{this} technique can be useful \edit{if either the characteristic error rate $\er$ or the number of qubits $n$ remains small (e.g. $n \er$ does not grow too large - see Appendix~\ref{app:d})}. Future work may further generalize the bounds we have derived and elucidate the non-perturbative regimes for which the errors in our methods remain well bounded.

% The technique described in this note is relevant to some applications in which a small error is tolerable. For instance, quantum kernel methods typically require that an experimental estimator $\hat{K}$ for an $m\times m$ matrix $K$ by computing $p(0)$ for a set of $\mathcal{O}(m^2)$ distinct circuits. In this case, the quantity $||\hat{K} - K||_F$ is predictive of the accuracy of the associated kernel classifier, such that a bounded element-wise error $[\hat{K} - K]_{ij}$ introduced by truncated readout correction can still yield acceptable classifier performance.

\edit{
At the time that this work was introduced, ref. \cite{wang2021measurement} appeared.
They also employed series approximations for computing $\Ro\inv$ (as discussed in \cref{sec:pert_full})
but the implementation is otherwise unrelated to the technique described here.
}

\section{Acknowledgements}
We thank Achim Kempf for reviewing this manuscript. EP is partially supported through A Kempf's Google Faculty Award. EP and GP are partially supported by the DOE/HEP QuantISED program grant HEP Machine Learning and Optimization Go Quantum, identification number 0000240323. A.C.Y.L.\ is supported by the DOE HEP QuantISED grant KA2401032. This manuscript has been authored by Fermi Research Alliance, LLC under Contract No. DE-AC02-07CH11359 with the U.S. Department of Energy, Office of Science, Office of High Energy Physics.
\edit{This research used resources of the Oak Ridge Leadership Computing Facility, which is a DOE Office of Science User Facility supported under Contract DE-AC05-00OR22725.}
% % % % % % % % % % % % % % % % % % % % % % % % % % % % % % % % % % % % % % % % 

\appendix

\section{Proof of Proposition~\ref{lemma:1}}\label{app:a}

The construction for $\Ro$ given in \cref{eq:type1} is a tensor product of identical single-qubit response matrices, each of which prescribes a fixed probability for a relaxation event $p(0|1) = \er$ and disallows excitation ($p(1|0) = 0$). The outline of the proof is that disallowing excitations results in an $\Ro$ with a block structure such that projection operations commute with the matrix product for strictly upper triangular submatrices of $\Ro$ acting over indices with weight less than $w$. The tensor structure then allows direct computation of $r$, and therefore also of $r_T$. 

$Q$ is upper triangular, and therefore \Ro ~is also upper triangular. Given the tensor structure of \Ro, we have that for indices $i,j\in\{0,1\}^n$ in the upper triangular set the elements of \Ro ~are given element-wise by
\begin{equation}\label{eq:Rij}
    \Ro_{ij} = 
    \begin{cases}
         \er^{w(j) - w(i)}(1-q)^{w(i)}, & \text{for } w(i) < w(j)\\
         (1-\er)^{w(i)}, & \text{for } i=j\\
        0, & \text{else } \\
 \end{cases}
\end{equation}
where the term $\er^{w(j) - w(i)}(1-\er)^{w(i)}$ represents the probability of $|w(j) - w(i)|$ simultaneous relaxations times the probability of the remaining $w(j) - |w(j)-w(i)| = w(i)$ bits \textit{not} relaxing. We split $\Ro$ into a diagonal component $\Rd$ and a strictly upper triangular \edit{component
\begin{equation}
    \Ro_u = \sum_{k=0}^n \sum_{\ell = k+1}^n |k\rangle \langle \ell| \otimes B_{k\ell},
\end{equation}
which represents a block matrix partition of $\Ro$ into $\{B_{k\ell}\}$, each of which} contains all elements $(\Ro)_{ij}$ with $w(i)=k$ and $w(j)=\ell$ and has dimensions
\begin{equation}
    \text{dim}(B_{k\ell}) = \begin{pmatrix} n \\ k \end{pmatrix} \times \begin{pmatrix} n \\ \ell \end{pmatrix}.
\end{equation}
\edit{$\Ro_u$ is an $(n+1)\times (n+1)$ block matrix, and is strictly upper triangular with respect to this block structure. For any strictly upper triangular $m \times m$ matrix $A$, we have that $A^m = 0$ \cite{lutkepohl1997handbook} and by similar reasoning $\Ro_u^{n+1} = 0$. Therefore, the Neumann series expansion for $\Ro^{-1}$ converges in $n+1$ terms and we obtain}
\begin{align}\label{eq:sum_expansion}
   \Ro^{-1}  &= \left(\sum_{k=0}^n (-\Rd^{-1}\Ro_u)^k \right) \Rd^{-1}.
\end{align}
In computing $(\Ro_u)^k$ every column space over basis vectors of weight $w$ depends only on contributions of column spaces over basis vectors of weight less than $w$ (i.e. the columns to the left of the weight-$w$ subspace). Therefore, we further partition $\Ro_u$ into column spaces $\{\sum_{k + \ell\geq w} B_{k\ell}\}_{w=1}^{n}$ with basis vectors less than or equal to $w$ and ignore the complementary column space for computing the truncated part of $R^{-1}$. 
That is, defining \edit{the projector} $\pi \equiv P_{0,w}$ and letting $\pi_\perp = I - \pi$ be the projector onto the complementary subspace of $\pi$ we simplify our representation of $\Ro_u$ as
\begin{equation}
\Ro_u \equiv \begin{pmatrix}
    \pi \\
    \pi_\perp
\end{pmatrix} 
\Ro_u 
\begin{pmatrix}
    \pi &    \pi_\perp
\end{pmatrix} 
\rightarrow 
\begin{pmatrix}
    \pi \Ro_u \pi & * \\
    0 & *
\end{pmatrix}
\end{equation}

The sum in \cref{eq:sum_expansion} may then be computed ignoring the columnspace of weights greater than $w$:
\begin{align}
&    (-\Rd^{-1}\Ro_u)^k \Rd^{-1}
\\
= & \label{eq:sum_term} 
 \begin{pmatrix}
    (-\pi \Rd^{-1} \Ro_u \pi)^k & * \\
    0 & *
\end{pmatrix}
  \begin{pmatrix}
    \pi \Rd^{-1} \pi & * \\
    0 & *
\end{pmatrix} 
\\
= & \begin{pmatrix}
    (-\pi \Rd^{-1} \Ro_u \pi)^k \pi \Rd^{-1} \pi  & * \\
    0 & *
\end{pmatrix}
\end{align}
since \edit{$\pi^2 = \pi$} by definition. Therefore,
\begin{align}
\pi (R^{-1}) \pi &= \pi \left( \sum_{k=0}^n (-\Rd^{-1}\Ro_u)^k \Rd^{-1} \right) \pi 
\\&= \sum_{k=0}^w (-\pi \Rd^{-1} \Ro_u \pi)^k (\pi \Rd^{-1} \pi) 
\\\label{eq:final}&= \sum_{k=0}^w \left(-\pi \Rd^{-1} \pi )(\pi \Ro_u \pi\right)^k (\pi \Rd^{-1} \pi) 
\\&\equiv (\pi R \pi )^{-1}
\end{align}
where the series now terminates at $w$ due to the nilpotency of $\pi \Ro_u \pi$ in \cref{eq:sum_term}. Line~\ref{eq:final} is simply the series expansion for the inverse of $\pi R \pi =\pi \Rd \pi + \pi \Ro_u \pi$, which concludes the proof.

% Now we apply the same treatement to compute the inverse of the truncated $R$. Since the columnspace of $\pi_{w}R =\pi_{w} \Rd + \pi_{w} \Ro_u$ has support only on 

% where $\Rd_w$ is diagonal and $T_u$ strictly upper triangular. Furthermore, the diagonal elements of $\Rd_w$  and apply the same treatment to compute the inverse of the truncated $R$:
% \begin{equation}
%     T^{-1} = \left(\sum_{k=0}^w (-S^{-1}T_u)^k \right) S^{-1}
% \end{equation}

\section{Proof of Theorem~\ref{thm:1} and Corollary~\ref{corollary:2}}\label{app:b}

The size of the projected subspace that includes all strings of weight less than or equal to $w$ is the binomial sum
\begin{equation}
    t(w) = \sum_{k=0}^w 
    \begin{pmatrix}
    n \\ k
\end{pmatrix}
\end{equation}
We can compute the error in the projected readout error correction method directly:
\begin{align}
      |  r_T \cdot p_T' -  r \cdot p'|   &= | (\left( \pi R \pi \right)^{-1} p_T')_0 -  (\Ro^{-1}p')_0| 
      \nonumber
      \\&= | (\pi \left( \Ro^{-1}  \right)\pi p_T')_0 -  (\Ro^{-1}p')_0| 
      \nonumber
      \\&=\left| \sum_{j=0}^{t(w)} (\Ro^{-1})_{0j} p_j' - \sum_{j=0}^{2^n-1} (\Ro^{-1})_{0j} p_j '  \right|
      \nonumber
      \\&= \left|  \sum_{j={t(w)}+1}^{2^n-1} (\Ro^{-1})_{0j} p_j '  \right|
      \nonumber
      \\ \label{eq:xx}&= \left|  \sum_{k = w+ 1}^{n} \left(\frac{\er}{1-\er}\right)^k \sum_{\ell \in B(w)} p_\ell '  \right| 
      \\ &\leq \left(\frac{\er}{1-\er}\right)^{w+1}
      \\&\leq \left(2\er\right)^{w+1}
\end{align}
where $B(w)$ denotes the set of bitstrings of weight $w$,
\edit{and $\pi \equiv P_{0,w}$ (as defined in Appendix \ref{app:a})}.
In line~\ref{eq:xx} the additional factor of $(1 - \er)^k$ is found by explicitly computing the first row of $\Ro^{-1}$: For a binary string $j = j_1 j_2\dots j_n$ the $j$-th entry of $r$ is:
\begin{align}
    |r_j| &= \left|(Q^{-1})_{0j_1} (Q^{-1})_{0j_2} \dots (Q^{-1})_{0j_n} \right|
    \\&= \left(\frac{\er}{(1-\er)}\right)^{w(j)}
\end{align}
Then with the requirement that $\er < 0.5$ we have $(1 - \er)^{-k} < 2^{-k}$ which completes the proof. To prove Corollary~\ref{corollary:2} we again explicitly compute $r_j$ taking advantage of the tensor structure of $\Ro^{-1}$:
\begin{align}
    |r_j| &= \left|(Q_1^{-1})_{0j_1} (Q_2^{-1})_{0j_2} \dots (Q_k^{-1})_{0j_n} \right|
    \\\label{eq:type2_r}&= \prod_{j_k}\left(\frac{\er_k}{(1-\er_k)}\right)^{j_k}
    \\&\leq \left(\max_k \frac{\er_k}{(1-\er_k)}\right)^{w(j)}
\end{align}
which upper bounds the magnitude of any element in the subspace of $\Ro^{-1}$ excluded by truncation of all bitstrings $j$ with $w(j) > w$. Proof of Corollary~\ref{corollary:2} proceeds identically as with Theorem~\ref{thm:1}. Note that the bound given is quite loose, and so the exact expression given in \cref{eq:type2_r} may be freely substituted if $\er_{max}$ is expected to be significantly larger than a typical $\er_k$.

% \begin{figure*}
%     \centering
%     \includegraphics[width=.65\linewidth]{figures/Rj_decomposition.pdf}
%     \caption{Decomposition of $\Ro$ into the set $\{\mathcal{R}_j\}_{j=0}^w$ initially yields sparse matrices each containing a fraction $s_j = 2^{-n} \big(\begin{smallmatrix}
%     n \\ j
%     \end{smallmatrix}\big)$ nonzero terms. As $w\rightarrow n/2$, the decomposition $\sum_{j=1}^w \mathcal{R}_{j}$ quickly becomes dense since $s_{n/2} \approx \frac{1}{2}$. Dashed lines indicate sectors of fixed-weight indices.}
%     \label{fig:Rj_decomposition}
% \end{figure*}

\section{Reduction of Algorithm \ref{alg:perturbative_mit_full} to single bitstrings}
\label{app:c}

Algorithm \ref{alg:perturbative_mit_full} can be further simplified if we are only interested in a specific element $\ell\in\{0,1\}^n$ from the distribution.
To illustrate the reduction, we construct a set $S_{\ell, w}$ consisting of all $M_{\ell, w} \equiv |S_{\ell, w}|$ basis states with at most a distance $w$ away from $\ell$:
\be
S_{\ell, w} = \{ \ m \ | \ s(m, \ell) \leq w \}.
\ee
We can define a projection operator $\Po_{\ell, w}$ given by
\be\label{eq:projection_operator}
\Po_{\ell, w} \hat{e}_n =
\begin{cases}
	\hat{e}_n,& \text{if } n \in S_{\ell, w}, \\
	0,              &  \text{otherwise}.
\end{cases}
\ee
It follows from \cref{eq:R_inverse_app} that the prior probability $\pideal_{\ell}$ for measuring a computational basis state $\sket{\ell}$ is given by
\begin{align}
\pideal_{\ell}
= &
\sum_{m}
\lf( \delta_{m, \ell}  + \sum_{k=1}^{w} \lf[ \lf(- \sum_{j=1}^{w} \er^{j} \Rd\inv \Rof{j} \rt)^k \rt]_{\ell,m} \rt)
\nn\\
& \times \lf(\Rd\rt)_{m,m}\inv \pexp_{m}
+ \mathcal{O} (\er^{w + 1})
.
\end{align}

It is straightforward to show that any matrix elements $[ (- \sum_{j=1}^{w} \er^{j} \Rd\inv \Rof{j} )^k ]_{\ell,m}$ with $s(m,\ell) > w$ are of order beyond $q^{w}$. We can thus use the projection operator $\Po_{\ell, w}$ to write
\begin{align}
& \lf[ \lf(- \sum_{j=1}^{w} \er^{j} \Rd\inv \Rof{j} \rt)^k \rt]_{\ell,m}
\nn\\
= &
\lf[\lf(- \sum_{j=1}^{w} \er^{j} \Po_{\ell, w} \Rd\inv \Po_{\ell, w}  \Rof{j} \Po_{\ell, w}  \rt)^k \rt]_{\ell,m} + \mathcal{O} (\er^{w + 1})
\nn
\end{align}
Defining the truncated operators to be $\RoTf{j}{\ell}{w} = \Po_{\ell, w}\Rof{j}\Po_{\ell, w}$, we get the approximated probability to be
\begin{align}
\papp_{\ell}
= &
\sum_{m}
\lf[ \delta_{m, \ell}  + \sum_{k=1}^{w} \lf(- \sum_{j=1}^{w} \er^{j} \lf(\RoTf{0}{\ell}{w}\rt)\inv \RoTf{j}{\ell}{w} \rt)^k  \rt]_{\ell,m}
\nn\\
&
\times \lf(\RoTf{0}{\ell}{w}\rt)_{m,m}
\inv \pexp_{m}
.
\end{align}
The algorithm to determine $\papp_{\ell}$ is similar to that for the full distribution but with $\Rof{j}$ being truncated to dimensions $M_{\ell, w} \times M_{\ell, w}$. The time complexity is thus given by $\mathcal{O}(M_{\ell, w}^3)$ which is consistent with the all-zeros bitstring case discussed in \Cref{sebsec:all-zeros_bitstring}.

Similarly to the case in \Cref{sec:pert_full}, the strategy of decomposing $\Ro$ into a set of sparse components $\{\mathcal{R}_j\}$ can be combined with alternatives to standard matrix inversion or series approximations to matrix inversion for recovering a specific bitstring $\ell$. For instance, the technique of \cite{Lee2014SolvingFA} for $\epsilon$-close approximation for specific elements of the solution to $Ax=b$ could be applied to recover $\pideal_\ell$ with exponential speedup over recovering the entire distribution $\pideal$ provided additional conditions on $\Ro$. However, the procedure to correct the readout probability for a specific bitstring $\ell\neq 0^n$ will generally require more resources than recovering the all-zeros bitstring. The subspace $S_{\ell, w}$ can be significantly larger than the subspace $S_{0, w}$, given that the majority of elements in $\{0,1\}^n$ have weight close to $\frac{n}{2}$. For example, to implement the algorithm of \Cref{sebsec:all-zeros_bitstring} to recover $\pideal_\ell$, $\Ro$ must be projected onto a subspace $\{x: w(\ell) - w_{min} \leq w(x) \leq w(\ell) + w_{max}\}$ consisting of strings with weight in $[w_{min}, w_{max}]$. This differs from the $\ell=0^n$ case since the population $p_0'$ cannot be increased due to the excitation of other bitstrings $x \neq 0$. 

From the perspective of readout error mitigation, computing $p_0$ (or $p(1^n)$, if necessary) is ideal, as it is the string with the fewest  neighbors separated by a low-weight error event. Consequently, if one desires to compute the probability of a fixed bitstring $\ell=\ell_1\ell_2\dots\ell_n$ at the output of a quantum circuit $U$, from the perspective of mitigating readout error it is preferable to perform readout rebalancing \cite{hicks_readout_2021} by appending a single layer of gates to construct  $U' = (\sigma_x^{\ell_1}\otimes \sigma_x^{\ell_2}\otimes \dots \otimes \sigma_x^{\ell_n})U$ where $\sigma_x$ is the pauli-X gate . Then the corrected probability $\papp_0$ sampled from the output of $U'$ is a maximally efficient approximation to $\pideal_\ell$ sampled from $U$.

\section{Modified perturbative technique}\label{app:d}

\begin{figure}
    \centering
    \includegraphics[width=\linewidth]{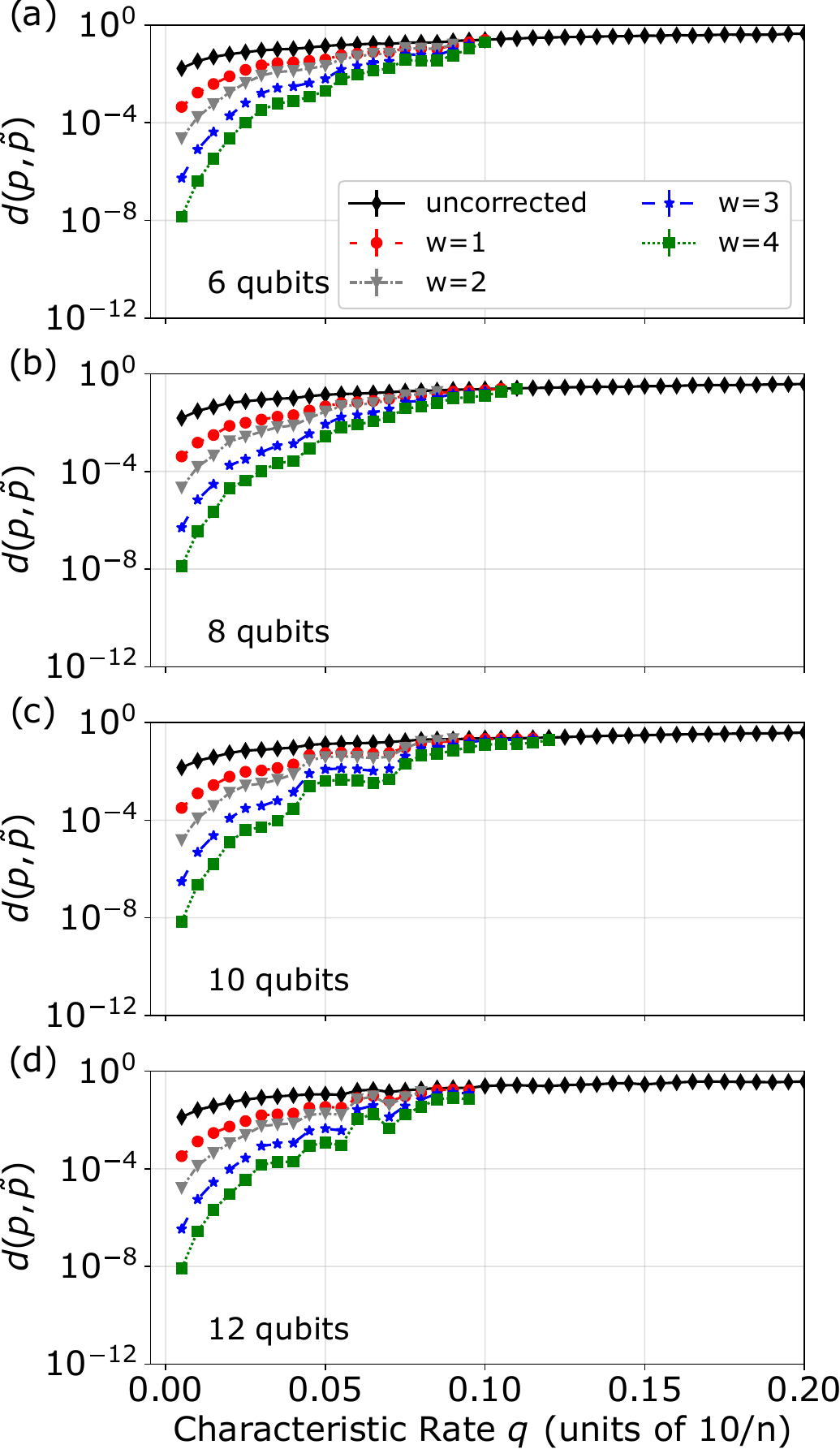}
    \caption{Failure points for variable number of qubits qubits occur around the same threshold $\er_{max}$ in units of $10 \frac{\er}{n}$. We empirically observe that this failure threshold scales inversely with $n$, signifying that our technique loses effectiveness when the characteristic error rate $\er$ cannot be suppressed as additional qubits are added to the system.}
    \label{fig:truncated_qstudy}
\end{figure}

As mentioned in the main text, our technique relies on the assumption that the Neumann series  for $\Ro\inv$ converges, namely
\begin{equation}\label{eq:norm_requirement}
    \norm{\sum_{j=1}^w \Rd\inv \mathcal{R}_j} < 1
\end{equation}
\edit{In general, increasing the number of qubits $n$ will uniformly increase the left-hand side of~\cref{eq:norm_requirement}, as the magnitude of the elements of each matrix $\Rof{j}$ is constant with respect to $n$ but the size of the matrix grows exponentially in $n$. Unless the characteristic error rate $q$ is reduced simultaneously as $n$ is increased (for example by bounding the product $nq$), the perturbative approximation of Algorithm~\ref{alg:perturbative_mit_full} will no longer hold}. \Cref{fig:truncated_qstudy} shows the effect of increasing $\er$ on the performance of Algorithm~\ref{alg:perturbative_mit_full}. The failure point of each experiment occurs when the error in $\papp$ is comparable to the error in $\pexp$, which we observed to typically coincide with reaching a characteristic rate $\er_{max}$ for which \cref{eq:norm_requirement} was no longer satisfied.

\begin{figure}
    \centering
    \includegraphics[width=\linewidth]{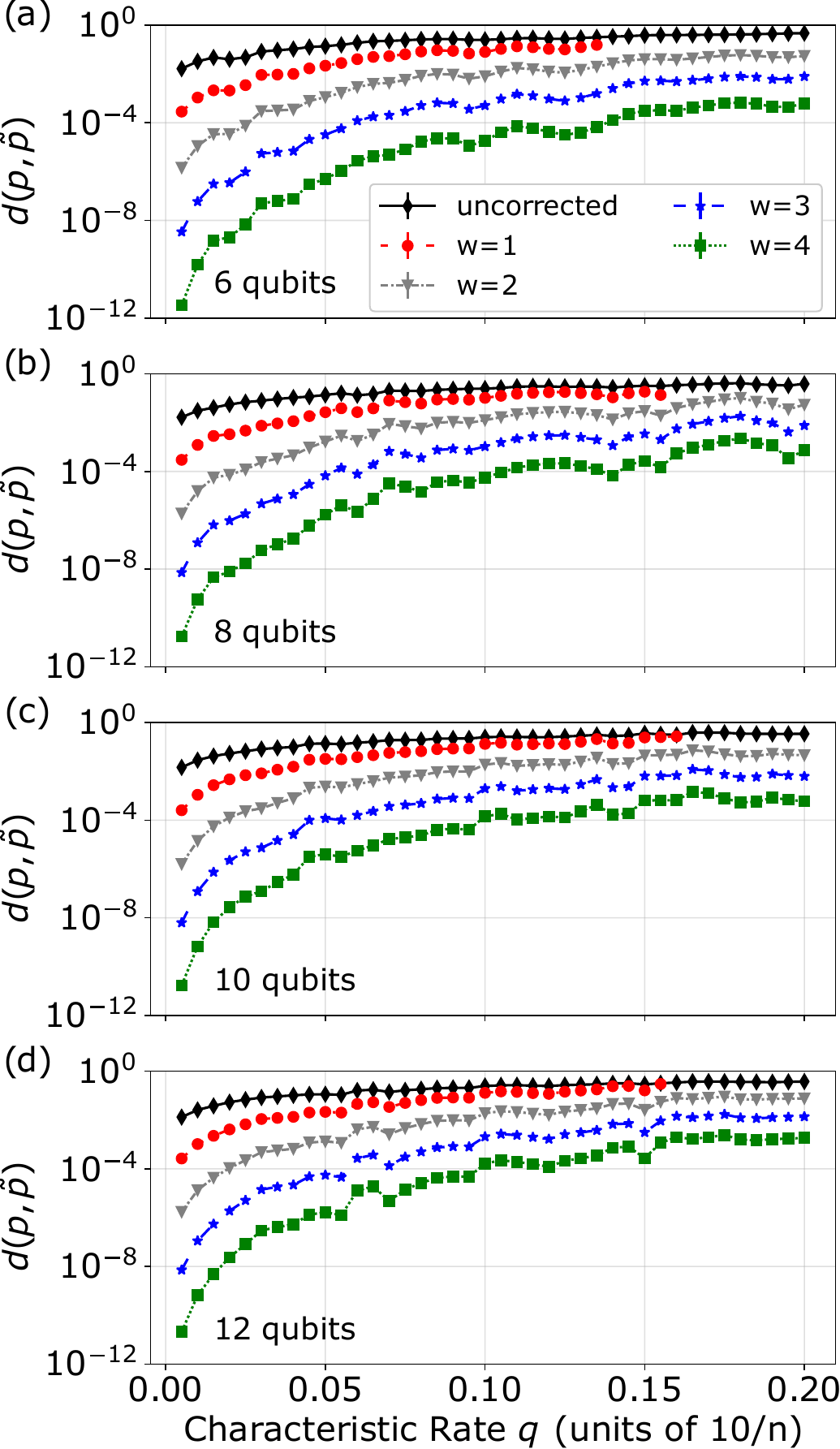}
    \caption{Our technique can be modified to overcome the limitations shown in \cref{fig:truncated_qstudy} by exactly inverting $\sum_{j=1}^w \mathcal{R}_j$. By doing so, the failure threshold $\er_{max}$ approaches $0.5$, indicating that the technique will work for arbitrary $\Ro$ constructed according to the model we have employed.}
    \label{fig:fullyinv_qstudy}
\end{figure}

Our technique may be modified slightly so that it still performs well even when the requirement of \cref{eq:norm_requirement} is no longer satisfied, since $\sum_{j=1}^w \mathcal{R}_j$ may still  invertible even when the Neumann series for its inverse does not converge. \Cref{fig:fullyinv_qstudy} shows the performance of this modified algorithm, and we highlight the fact that this performance improves steadily with increasing $w$. This improvement comes at the cost of a larger classical computation overhead, but this scenario may still be preferable over completely characterizing $\Ro$ with an exponentially large diagnostic experiment.

\begin{figure}
	\centering
	\includegraphics[width=\linewidth ]{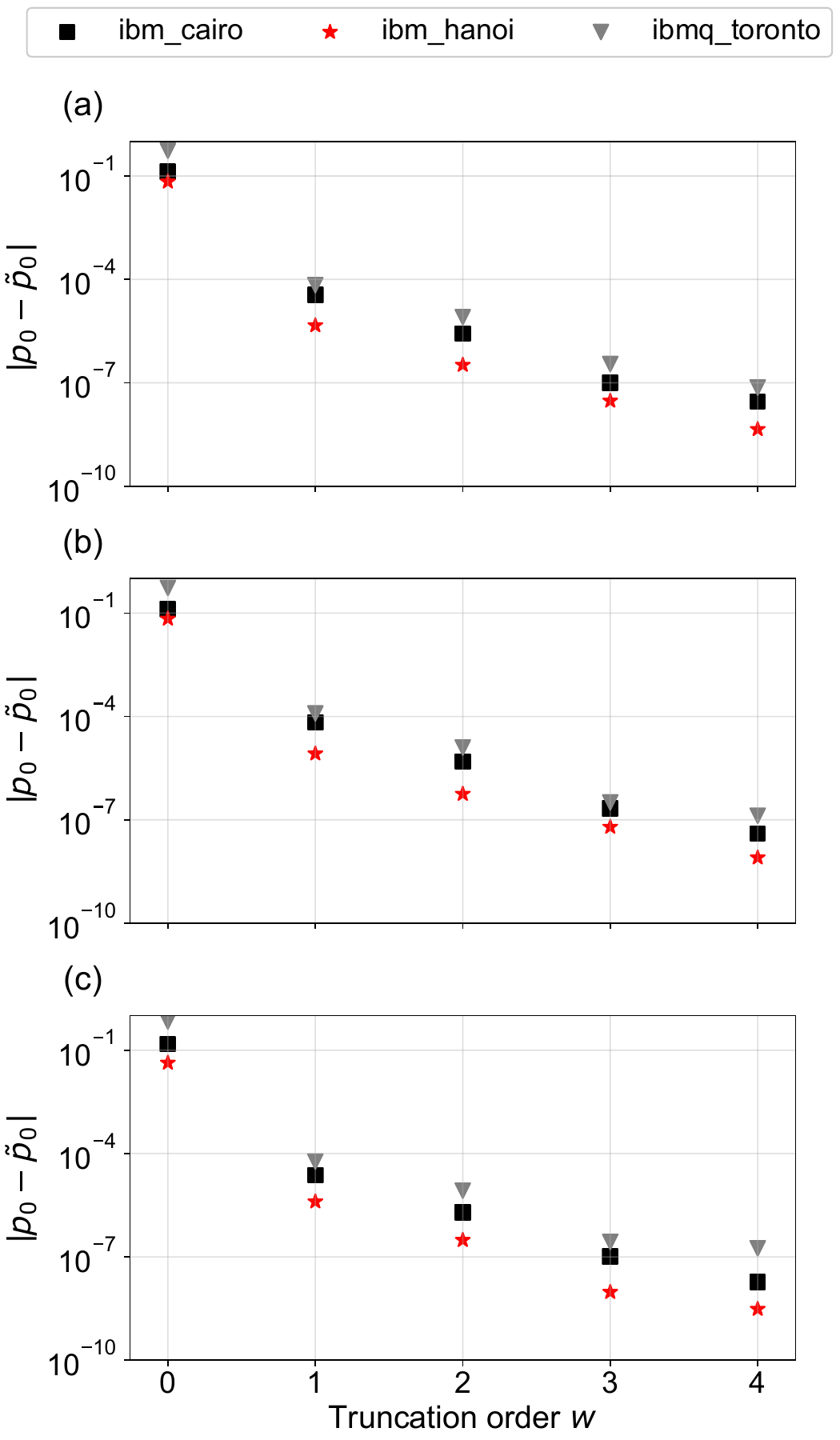}
	\caption{
		\editandy{
			Performance of all-zeros readout error mitigation with response matrices experimentally determined on IBM QPUs. We demonstrate the performance using three prior distributions, namely, (a) Gaussian distribution, (b) truncated Gaussian distribution and (c) uniform distribution. The details of the distributions are explained in the caption of \cref{fig:numeric1}.
		}
	}
	\label{fig:ibmq_results}
\end{figure}

\editandy{
    \section{Additional numerics}
	\subsection{Response matrices measured on IBM QPUs}
	\label{app:numerics_QPU_results}
}

\editandy{
	Our numerical experiments in the main text used response matrices generated with a tensor-structure assumption. In reality, the response matrix does not have a tensor structure in general, though the tensor structure could be a good approximation in many cases. We now implement our technique using response matrices measured experimentally on IBM QPUs and demonstrate that the efficacy of the technique for realistic readout errors on NISQ devices.
}

\editandy{
	We measured the response matrices on three different 27-qubit IBM QPUs, namely 'ibm\_cairo', 'ibm\_hanoi' and 'ibmq\_toronto'.
	The response matrices were estimated by preparing computational basis states $| j_1 \dots j_n)$ using a sequence of X gates, i.e., $X_1^{j_1} \dots X_n^{j_n}$, and then determining the output distributions via parallel qubit readout. Each of these measurements requires $2^n$ different circuit executions with each execution to be repeated several times to generate a distribution. To avoid exponential resource overhead, we estimated $\Ro$ for 12 out of 27 qubits using 10000 shots per matrix element. In particular, we picked qubits 1, 2, 3, 5, 8, 11, 14, 16, 19, 22, 25 and 26 for our experiment. Note that all three QPUs have the same connectivity map.
}

\editandy{
	\cref{fig:ibmq_results} shows the performance of our technique for recovering the all-zeros bitstrings using the response matrices measured on IBM hardware. Similar to the result using a tensor-structure assumption shown in \cref{fig:numeric1}, the error is exponentially suppressed in the truncation order $w$.
}

\begin{figure}
    \centering
    \includegraphics[width=\linewidth ]{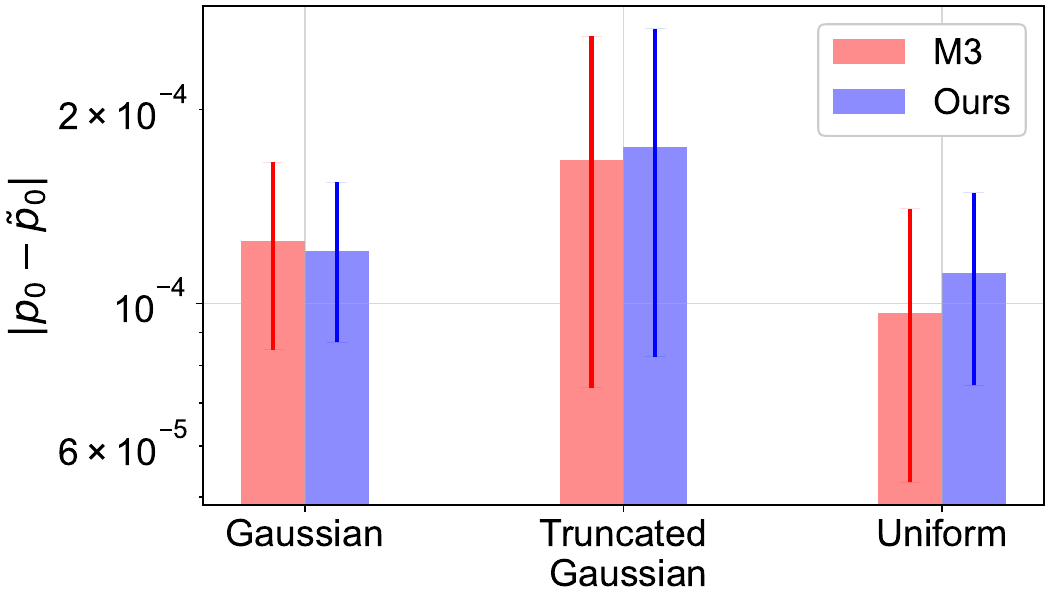}
    \caption{\edit{The accuracy of our technique for recovering the all-zeros bitstring is statistically indistinguishable to that of M3 for the distributions considered in \cref{fig:numeric1} using $n=8$ qubits, with $M=N=10^6$ circuit repetitions used for both the calibration experiment and for sampling $p'$ (see main text). The response matrix $\Ro$ has the same characteristic error rate $q$ as in previous numerics. Error bars denote standard deviation.}}
    \label{fig:m3}
\end{figure}

\edit{
\subsection{Sampling error and comparison to 'M3'}\label{app:m3}
In this section, we provide additional numerical experiments comparing the performance of our method to existing methods. We provide preliminary evidence that our technique for estimating the all-zeros bitstring provides comparable accuracy as the `M3' technique of Ref.~\cite{PRXQuantum.2.040326} for instances tested on $8$ qubits. M3 performs readout error correction by operating in a subspace corresponding to bitstrings that were sampled in an experiment with finite repetitions. Thus, M3 corrects an empirical distribution $h' \in \mathbb{R}^{2^n}$ sampled according to the observed bitstring probability vector $p'$ and takes as input a response matrix $\Ro$ sampled from calibration circuits on hardware. To introduce similar sampling error into our technique, we prepared independent qubit bitflip probabilities $\epsilon_k, \eta_k \sim \text{Uniform}(0, q)$ and then prepared estimates $\tilde{\epsilon}_k := B(N, \epsilon_k)/N$, $\tilde{\eta}_k := B(N, \eta_k)/ N$ (where $B$ is the binomial distribution) to simulate a series of independent calibration experiments using $N$ circuit repetitions for each of qubits $k=1, \dots, n$. We then used a sampled response matrix
\begin{equation}
    \tilde{\Ro} = \bigotimes_{k=1}^n \begin{pmatrix}
    1 - \tilde{\eta}_k & \tilde{\epsilon}_k \\ 
    \tilde{\epsilon}_k & 1 - \tilde{\epsilon}_k
\end{pmatrix}
\end{equation}
in place of $\Ro$ for all parts of our algorithm. Similarly, we substituted an empirical estimate $h'$ with components $h_i' = B(M, p_i')/M$ for the probability vector over observed bitstrings $p'$ to simulate sampling a circuit run for $M$ repetitions. 
In \cref{fig:m3}, we numerically simulate recovering the all-zeros bitstring for the eight-qubit case using our technique and the M3 technique. The simulation shows that the two techniques give results with a similar accuracy.
}

% \newpage
% \bibliographystyle{apsrev4-1}
% \bibliography{references}

\begin{thebibliography}{0}%
\makeatletter
\providecommand \@ifxundefined [1]{%
 \@ifx{#1\undefined}
}%
\providecommand \@ifnum [1]{%
 \ifnum #1\expandafter \@firstoftwo
 \else \expandafter \@secondoftwo
 \fi
}%
\providecommand \@ifx [1]{%
 \ifx #1\expandafter \@firstoftwo
 \else \expandafter \@secondoftwo
 \fi
}%
\providecommand \natexlab [1]{#1}%
\providecommand \enquote  [1]{``#1''}%
\providecommand \bibnamefont  [1]{#1}%
\providecommand \bibfnamefont [1]{#1}%
\providecommand \citenamefont [1]{#1}%
\providecommand \href@noop [0]{\@secondoftwo}%
\providecommand \href [0]{\begingroup \@sanitize@url \@href}%
\providecommand \@href[1]{\@@startlink{#1}\@@href}%
\providecommand \@@href[1]{\endgroup#1\@@endlink}%
\providecommand \@sanitize@url [0]{\catcode `\\12\catcode `\$12\catcode
  `\&12\catcode `\#12\catcode `\^12\catcode `\_12\catcode `\%12\relax}%
\providecommand \@@startlink[1]{}%
\providecommand \@@endlink[0]{}%
\providecommand \url  [0]{\begingroup\@sanitize@url \@url }%
\providecommand \@url [1]{\endgroup\@href {#1}{\urlprefix }}%
\providecommand \urlprefix  [0]{URL }%
\providecommand \Eprint [0]{\href }%
\providecommand \doibase [0]{https://doi.org/}%
\providecommand \selectlanguage [0]{\@gobble}%
\providecommand \bibinfo  [0]{\@secondoftwo}%
\providecommand \bibfield  [0]{\@secondoftwo}%
\providecommand \translation [1]{[#1]}%
\providecommand \BibitemOpen [0]{}%
\providecommand \bibitemStop [0]{}%
\providecommand \bibitemNoStop [0]{.\EOS\space}%
\providecommand \EOS [0]{\spacefactor3000\relax}%
\providecommand \BibitemShut  [1]{\csname bibitem#1\endcsname}%
\let\auto@bib@innerbib\@empty
%</preamble>
\end{thebibliography}%


\begin{thebibliography}{31}%
\makeatletter
\providecommand \@ifxundefined [1]{%
 \@ifx{#1\undefined}
}%
\providecommand \@ifnum [1]{%
 \ifnum #1\expandafter \@firstoftwo
 \else \expandafter \@secondoftwo
 \fi
}%
\providecommand \@ifx [1]{%
 \ifx #1\expandafter \@firstoftwo
 \else \expandafter \@secondoftwo
 \fi
}%
\providecommand \natexlab [1]{#1}%
\providecommand \enquote  [1]{``#1''}%
\providecommand \bibnamefont  [1]{#1}%
\providecommand \bibfnamefont [1]{#1}%
\providecommand \citenamefont [1]{#1}%
\providecommand \href@noop [0]{\@secondoftwo}%
\providecommand \href [0]{\begingroup \@sanitize@url \@href}%
\providecommand \@href[1]{\@@startlink{#1}\@@href}%
\providecommand \@@href[1]{\endgroup#1\@@endlink}%
\providecommand \@sanitize@url [0]{\catcode `\\12\catcode `\$12\catcode
  `\&12\catcode `\#12\catcode `\^12\catcode `\_12\catcode `\%12\relax}%
\providecommand \@@startlink[1]{}%
\providecommand \@@endlink[0]{}%
\providecommand \url  [0]{\begingroup\@sanitize@url \@url }%
\providecommand \@url [1]{\endgroup\@href {#1}{\urlprefix }}%
\providecommand \urlprefix  [0]{URL }%
\providecommand \Eprint [0]{\href }%
\providecommand \doibase [0]{http://dx.doi.org/}%
\providecommand \selectlanguage [0]{\@gobble}%
\providecommand \bibinfo  [0]{\@secondoftwo}%
\providecommand \bibfield  [0]{\@secondoftwo}%
\providecommand \translation [1]{[#1]}%
\providecommand \BibitemOpen [0]{}%
\providecommand \bibitemStop [0]{}%
\providecommand \bibitemNoStop [0]{.\EOS\space}%
\providecommand \EOS [0]{\spacefactor3000\relax}%
\providecommand \BibitemShut  [1]{\csname bibitem#1\endcsname}%
\let\auto@bib@innerbib\@empty
%</preamble>
\bibitem [{\citenamefont {Preskill}(2018)}]{Preskill2018quantumcomputingin}%
  \BibitemOpen
  \bibfield  {author} {\bibinfo {author} {\bibfnamefont {J.}~\bibnamefont
  {Preskill}},\ }\href {\doibase 10.22331/q-2018-08-06-79} {\bibfield
  {journal} {\bibinfo  {journal} {{Quantum}}\ }\textbf {\bibinfo {volume}
  {2}},\ \bibinfo {pages} {79} (\bibinfo {year} {2018})}\BibitemShut {NoStop}%
\bibitem [{\citenamefont {Neeley}\ \emph {et~al.}(2010)\citenamefont {Neeley},
  \citenamefont {Bialczak}, \citenamefont {Lenander}, \citenamefont {Lucero},
  \citenamefont {Mariantoni}, \citenamefont {O’Connell}, \citenamefont
  {Sank}, \citenamefont {Wang}, \citenamefont {Weides}, \citenamefont {Wenner}
  \emph {et~al.}}]{Neeley_2010}%
  \BibitemOpen
  \bibfield  {author} {\bibinfo {author} {\bibfnamefont {M.}~\bibnamefont
  {Neeley}}, \bibinfo {author} {\bibfnamefont {R.~C.}\ \bibnamefont
  {Bialczak}}, \bibinfo {author} {\bibfnamefont {M.}~\bibnamefont {Lenander}},
  \bibinfo {author} {\bibfnamefont {E.}~\bibnamefont {Lucero}}, \bibinfo
  {author} {\bibfnamefont {M.}~\bibnamefont {Mariantoni}}, \bibinfo {author}
  {\bibfnamefont {A.~D.}\ \bibnamefont {O’Connell}}, \bibinfo {author}
  {\bibfnamefont {D.}~\bibnamefont {Sank}}, \bibinfo {author} {\bibfnamefont
  {H.}~\bibnamefont {Wang}}, \bibinfo {author} {\bibfnamefont {M.}~\bibnamefont
  {Weides}}, \bibinfo {author} {\bibfnamefont {J.}~\bibnamefont {Wenner}},
  \emph {et~al.},\ }\href {\doibase 10.1038/nature09418} {\bibfield  {journal}
  {\bibinfo  {journal} {Nature}\ }\textbf {\bibinfo {volume} {467}},\ \bibinfo
  {pages} {570–573} (\bibinfo {year} {2010})}\BibitemShut {NoStop}%
\bibitem [{\citenamefont {Willsch}\ \emph {et~al.}(2018)\citenamefont
  {Willsch}, \citenamefont {Willsch}, \citenamefont {Jin}, \citenamefont
  {De~Raedt},\ and\ \citenamefont {Michielsen}}]{Willsch_2018}%
  \BibitemOpen
  \bibfield  {author} {\bibinfo {author} {\bibfnamefont {D.}~\bibnamefont
  {Willsch}}, \bibinfo {author} {\bibfnamefont {M.}~\bibnamefont {Willsch}},
  \bibinfo {author} {\bibfnamefont {F.}~\bibnamefont {Jin}}, \bibinfo {author}
  {\bibfnamefont {H.}~\bibnamefont {De~Raedt}}, \ and\ \bibinfo {author}
  {\bibfnamefont {K.}~\bibnamefont {Michielsen}},\ }\href {\doibase
  10.1103/physreva.98.052348} {\bibfield  {journal} {\bibinfo  {journal} {Phys.
  Rev. A}\ }\textbf {\bibinfo {volume} {98}} (\bibinfo {year} {2018}),\
  10.1103/physreva.98.052348}\BibitemShut {NoStop}%
\bibitem [{\citenamefont {Dewes}\ \emph {et~al.}(2012)\citenamefont {Dewes},
  \citenamefont {Ong}, \citenamefont {Schmitt}, \citenamefont {Lauro},
  \citenamefont {Boulant}, \citenamefont {Bertet}, \citenamefont {Vion},\ and\
  \citenamefont {Esteve}}]{Dewes_2012}%
  \BibitemOpen
  \bibfield  {author} {\bibinfo {author} {\bibfnamefont {A.}~\bibnamefont
  {Dewes}}, \bibinfo {author} {\bibfnamefont {F.~R.}\ \bibnamefont {Ong}},
  \bibinfo {author} {\bibfnamefont {V.}~\bibnamefont {Schmitt}}, \bibinfo
  {author} {\bibfnamefont {R.}~\bibnamefont {Lauro}}, \bibinfo {author}
  {\bibfnamefont {N.}~\bibnamefont {Boulant}}, \bibinfo {author} {\bibfnamefont
  {P.}~\bibnamefont {Bertet}}, \bibinfo {author} {\bibfnamefont
  {D.}~\bibnamefont {Vion}}, \ and\ \bibinfo {author} {\bibfnamefont
  {D.}~\bibnamefont {Esteve}},\ }\href {\doibase
  10.1103/physrevlett.108.057002} {\bibfield  {journal} {\bibinfo  {journal}
  {Physical Review Letters}\ }\textbf {\bibinfo {volume} {108}} (\bibinfo
  {year} {2012}),\ 10.1103/physrevlett.108.057002}\BibitemShut {NoStop}%
\bibitem [{\citenamefont {Chen}\ \emph {et~al.}(2019)\citenamefont {Chen},
  \citenamefont {Farahzad}, \citenamefont {Yoo},\ and\ \citenamefont
  {Wei}}]{Chen_2019}%
  \BibitemOpen
  \bibfield  {author} {\bibinfo {author} {\bibfnamefont {Y.}~\bibnamefont
  {Chen}}, \bibinfo {author} {\bibfnamefont {M.}~\bibnamefont {Farahzad}},
  \bibinfo {author} {\bibfnamefont {S.}~\bibnamefont {Yoo}}, \ and\ \bibinfo
  {author} {\bibfnamefont {T.-C.}\ \bibnamefont {Wei}},\ }\href {\doibase
  10.1103/physreva.100.052315} {\bibfield  {journal} {\bibinfo  {journal}
  {Phys. Rev. A}\ }\textbf {\bibinfo {volume} {100}} (\bibinfo {year} {2019}),\
  10.1103/physreva.100.052315}\BibitemShut {NoStop}%
\bibitem [{\citenamefont {Gong}\ \emph {et~al.}(2019)\citenamefont {Gong},
  \citenamefont {Chen}, \citenamefont {Zheng}, \citenamefont {Wang},
  \citenamefont {Zha}, \citenamefont {Deng}, \citenamefont {Yan}, \citenamefont
  {Rong}, \citenamefont {Wu}, \citenamefont {Li},\ and\ \citenamefont
  {et~al.}}]{Gong_2019}%
  \BibitemOpen
  \bibfield  {author} {\bibinfo {author} {\bibfnamefont {M.}~\bibnamefont
  {Gong}}, \bibinfo {author} {\bibfnamefont {M.-C.}\ \bibnamefont {Chen}},
  \bibinfo {author} {\bibfnamefont {Y.}~\bibnamefont {Zheng}}, \bibinfo
  {author} {\bibfnamefont {S.}~\bibnamefont {Wang}}, \bibinfo {author}
  {\bibfnamefont {C.}~\bibnamefont {Zha}}, \bibinfo {author} {\bibfnamefont
  {H.}~\bibnamefont {Deng}}, \bibinfo {author} {\bibfnamefont {Z.}~\bibnamefont
  {Yan}}, \bibinfo {author} {\bibfnamefont {H.}~\bibnamefont {Rong}}, \bibinfo
  {author} {\bibfnamefont {Y.}~\bibnamefont {Wu}}, \bibinfo {author}
  {\bibfnamefont {S.}~\bibnamefont {Li}}, \ and\ \bibinfo {author}
  {\bibnamefont {et~al.}},\ }\href {\doibase 10.1103/physrevlett.122.110501}
  {\bibfield  {journal} {\bibinfo  {journal} {Phys. Rev. Lett.}\ }\textbf
  {\bibinfo {volume} {122}} (\bibinfo {year} {2019}),\
  10.1103/physrevlett.122.110501}\BibitemShut {NoStop}%
\bibitem [{\citenamefont {Havl{\'{\i}}{\v{c}}ek}\ \emph
  {et~al.}(2019)\citenamefont {Havl{\'{\i}}{\v{c}}ek}, \citenamefont
  {C{\'{o}}rcoles}, \citenamefont {Temme}, \citenamefont {Harrow},
  \citenamefont {Kandala}, \citenamefont {Chow},\ and\ \citenamefont
  {Gambetta}}]{havlicek_2019}%
  \BibitemOpen
  \bibfield  {author} {\bibinfo {author} {\bibfnamefont {V.}~\bibnamefont
  {Havl{\'{\i}}{\v{c}}ek}}, \bibinfo {author} {\bibfnamefont {A.~D.}\
  \bibnamefont {C{\'{o}}rcoles}}, \bibinfo {author} {\bibfnamefont
  {K.}~\bibnamefont {Temme}}, \bibinfo {author} {\bibfnamefont {A.~W.}\
  \bibnamefont {Harrow}}, \bibinfo {author} {\bibfnamefont {A.}~\bibnamefont
  {Kandala}}, \bibinfo {author} {\bibfnamefont {J.~M.}\ \bibnamefont {Chow}}, \
  and\ \bibinfo {author} {\bibfnamefont {J.~M.}\ \bibnamefont {Gambetta}},\
  }\href {\doibase 10.1038/s41586-019-0980-2} {\bibfield  {journal} {\bibinfo
  {journal} {Nature}\ }\textbf {\bibinfo {volume} {567}},\ \bibinfo {pages}
  {209} (\bibinfo {year} {2019})}\BibitemShut {NoStop}%
\bibitem [{\citenamefont {Wei}\ \emph {et~al.}(2020)\citenamefont {Wei},
  \citenamefont {Lauer}, \citenamefont {Srinivasan}, \citenamefont
  {Sundaresan}, \citenamefont {McClure}, \citenamefont {Toyli}, \citenamefont
  {McKay}, \citenamefont {Gambetta},\ and\ \citenamefont {Sheldon}}]{Wei_2020}%
  \BibitemOpen
  \bibfield  {author} {\bibinfo {author} {\bibfnamefont {K.~X.}\ \bibnamefont
  {Wei}}, \bibinfo {author} {\bibfnamefont {I.}~\bibnamefont {Lauer}}, \bibinfo
  {author} {\bibfnamefont {S.}~\bibnamefont {Srinivasan}}, \bibinfo {author}
  {\bibfnamefont {N.}~\bibnamefont {Sundaresan}}, \bibinfo {author}
  {\bibfnamefont {D.~T.}\ \bibnamefont {McClure}}, \bibinfo {author}
  {\bibfnamefont {D.}~\bibnamefont {Toyli}}, \bibinfo {author} {\bibfnamefont
  {D.~C.}\ \bibnamefont {McKay}}, \bibinfo {author} {\bibfnamefont {J.~M.}\
  \bibnamefont {Gambetta}}, \ and\ \bibinfo {author} {\bibfnamefont
  {S.}~\bibnamefont {Sheldon}},\ }\href {\doibase 10.1103/physreva.101.032343}
  {\bibfield  {journal} {\bibinfo  {journal} {Phys. Rev. A}\ }\textbf {\bibinfo
  {volume} {101}} (\bibinfo {year} {2020}),\
  10.1103/physreva.101.032343}\BibitemShut {NoStop}%
\bibitem [{\citenamefont {Urbanek}\ \emph {et~al.}(2020)\citenamefont
  {Urbanek}, \citenamefont {Nachman},\ and\ \citenamefont
  {de~Jong}}]{Urbanek_2020}%
  \BibitemOpen
  \bibfield  {author} {\bibinfo {author} {\bibfnamefont {M.}~\bibnamefont
  {Urbanek}}, \bibinfo {author} {\bibfnamefont {B.}~\bibnamefont {Nachman}}, \
  and\ \bibinfo {author} {\bibfnamefont {W.~A.}\ \bibnamefont {de~Jong}},\
  }\href {\doibase 10.1103/physreva.102.022427} {\bibfield  {journal} {\bibinfo
   {journal} {Phys. Rev. A}\ }\textbf {\bibinfo {volume} {102}} (\bibinfo
  {year} {2020}),\ 10.1103/physreva.102.022427}\BibitemShut {NoStop}%
\bibitem [{\citenamefont {Nachman}\ \emph {et~al.}(2020)\citenamefont
  {Nachman}, \citenamefont {Urbanek}, \citenamefont {de~Jong},\ and\
  \citenamefont {Bauer}}]{nachman_unfolding_2020}%
  \BibitemOpen
  \bibfield  {author} {\bibinfo {author} {\bibfnamefont {B.}~\bibnamefont
  {Nachman}}, \bibinfo {author} {\bibfnamefont {M.}~\bibnamefont {Urbanek}},
  \bibinfo {author} {\bibfnamefont {W.~A.}\ \bibnamefont {de~Jong}}, \ and\
  \bibinfo {author} {\bibfnamefont {C.~W.}\ \bibnamefont {Bauer}},\ }\href
  {\doibase 10.1038/s41534-020-00309-7} {\bibfield  {journal} {\bibinfo
  {journal} {npj Quantum Inf.}\ }\textbf {\bibinfo {volume} {6}},\ \bibinfo
  {pages} {1} (\bibinfo {year} {2020})}\BibitemShut {NoStop}%
\bibitem [{\citenamefont {Hicks}\ \emph {et~al.}(2021)\citenamefont {Hicks},
  \citenamefont {Bauer},\ and\ \citenamefont {Nachman}}]{hicks_readout_2021}%
  \BibitemOpen
  \bibfield  {author} {\bibinfo {author} {\bibfnamefont {R.}~\bibnamefont
  {Hicks}}, \bibinfo {author} {\bibfnamefont {C.~W.}\ \bibnamefont {Bauer}}, \
  and\ \bibinfo {author} {\bibfnamefont {B.}~\bibnamefont {Nachman}},\ }\href
  {\doibase 10.1103/PhysRevA.103.022407} {\bibfield  {journal} {\bibinfo
  {journal} {Phys. Rev. A}\ }\textbf {\bibinfo {volume} {103}} (\bibinfo {year}
  {2021}),\ 10.1103/PhysRevA.103.022407}\BibitemShut {NoStop}%
\bibitem [{\citenamefont {Funcke}\ \emph {et~al.}(2022)\citenamefont {Funcke},
  \citenamefont {Hartung}, \citenamefont {Jansen}, \citenamefont {Kühn},
  \citenamefont {Stornati},\ and\ \citenamefont
  {Wang}}]{funcke2020measurement}%
  \BibitemOpen
  \bibfield  {author} {\bibinfo {author} {\bibfnamefont {L.}~\bibnamefont
  {Funcke}}, \bibinfo {author} {\bibfnamefont {T.}~\bibnamefont {Hartung}},
  \bibinfo {author} {\bibfnamefont {K.}~\bibnamefont {Jansen}}, \bibinfo
  {author} {\bibfnamefont {S.}~\bibnamefont {Kühn}}, \bibinfo {author}
  {\bibfnamefont {P.}~\bibnamefont {Stornati}}, \ and\ \bibinfo {author}
  {\bibfnamefont {X.}~\bibnamefont {Wang}},\ }\href {\doibase
  10.1103/physreva.105.062404} {\bibfield  {journal} {\bibinfo  {journal}
  {Phys. Rev. A}\ }\textbf {\bibinfo {volume} {105}} (\bibinfo {year} {2022}),\
  10.1103/physreva.105.062404}\BibitemShut {NoStop}%
\bibitem [{\citenamefont {Bravyi}\ \emph {et~al.}(2021)\citenamefont {Bravyi},
  \citenamefont {Sheldon}, \citenamefont {Kandala}, \citenamefont {Mckay},\
  and\ \citenamefont {Gambetta}}]{bravyi2020mitigating}%
  \BibitemOpen
  \bibfield  {author} {\bibinfo {author} {\bibfnamefont {S.}~\bibnamefont
  {Bravyi}}, \bibinfo {author} {\bibfnamefont {S.}~\bibnamefont {Sheldon}},
  \bibinfo {author} {\bibfnamefont {A.}~\bibnamefont {Kandala}}, \bibinfo
  {author} {\bibfnamefont {D.~C.}\ \bibnamefont {Mckay}}, \ and\ \bibinfo
  {author} {\bibfnamefont {J.~M.}\ \bibnamefont {Gambetta}},\ }\href {\doibase
  10.1103/physreva.103.042605} {\bibfield  {journal} {\bibinfo  {journal}
  {Phys. Rev. A}\ }\textbf {\bibinfo {volume} {103}} (\bibinfo {year} {2021}),\
  10.1103/physreva.103.042605}\BibitemShut {NoStop}%
\bibitem [{\citenamefont {van~den Berg}\ \emph {et~al.}(2022)\citenamefont
  {van~den Berg}, \citenamefont {Minev},\ and\ \citenamefont
  {Temme}}]{PhysRevA.105.032620}%
  \BibitemOpen
  \bibfield  {author} {\bibinfo {author} {\bibfnamefont {E.}~\bibnamefont
  {van~den Berg}}, \bibinfo {author} {\bibfnamefont {Z.~K.}\ \bibnamefont
  {Minev}}, \ and\ \bibinfo {author} {\bibfnamefont {K.}~\bibnamefont
  {Temme}},\ }\href {\doibase 10.1103/PhysRevA.105.032620} {\bibfield
  {journal} {\bibinfo  {journal} {Phys. Rev. A}\ }\textbf {\bibinfo {volume}
  {105}},\ \bibinfo {pages} {032620} (\bibinfo {year} {2022})}\BibitemShut
  {NoStop}%
\bibitem [{\citenamefont {Peters}\ \emph {et~al.}(2021)\citenamefont {Peters},
  \citenamefont {Caldeira}, \citenamefont {Ho}, \citenamefont {Leichenauer},
  \citenamefont {Mohseni}, \citenamefont {Neven}, \citenamefont {Spentzouris},
  \citenamefont {Strain},\ and\ \citenamefont {Perdue}}]{peters2021machine}%
  \BibitemOpen
  \bibfield  {author} {\bibinfo {author} {\bibfnamefont {E.}~\bibnamefont
  {Peters}}, \bibinfo {author} {\bibfnamefont {J.}~\bibnamefont {Caldeira}},
  \bibinfo {author} {\bibfnamefont {A.}~\bibnamefont {Ho}}, \bibinfo {author}
  {\bibfnamefont {S.}~\bibnamefont {Leichenauer}}, \bibinfo {author}
  {\bibfnamefont {M.}~\bibnamefont {Mohseni}}, \bibinfo {author} {\bibfnamefont
  {H.}~\bibnamefont {Neven}}, \bibinfo {author} {\bibfnamefont
  {P.}~\bibnamefont {Spentzouris}}, \bibinfo {author} {\bibfnamefont
  {D.}~\bibnamefont {Strain}}, \ and\ \bibinfo {author} {\bibfnamefont {G.~N.}\
  \bibnamefont {Perdue}},\ }\href {\doibase 10.1038/s41534-021-00498-9}
  {\bibfield  {journal} {\bibinfo  {journal} {npj Quantum Inf.}\ }\textbf
  {\bibinfo {volume} {7}},\ \bibinfo {pages} {1} (\bibinfo {year}
  {2021})}\BibitemShut {NoStop}%
\bibitem [{\citenamefont {Nachman}\ and\ \citenamefont
  {Geller}(2021)}]{nachman_2021}%
  \BibitemOpen
  \bibfield  {author} {\bibinfo {author} {\bibfnamefont {B.}~\bibnamefont
  {Nachman}}\ and\ \bibinfo {author} {\bibfnamefont {M.~R.}\ \bibnamefont
  {Geller}},\ }\href {\doibase 10.48550/ARXIV.2104.04607} {\enquote {\bibinfo
  {title} {Categorizing readout error correlations on near term quantum
  computers},}\ } (\bibinfo {year} {2021}),\ \Eprint
  {http://arxiv.org/abs/2104.04607} {arXiv:2104.04607 [quant-ph]} \BibitemShut
  {NoStop}%
\bibitem [{\citenamefont {Geller}(2020)}]{geller_rigorous_2020}%
  \BibitemOpen
  \bibfield  {author} {\bibinfo {author} {\bibfnamefont {M.~R.}\ \bibnamefont
  {Geller}},\ }\href {\doibase 10.1088/2058-9565/ab9591} {\bibfield  {journal}
  {\bibinfo  {journal} {Quantum Sci. Technol.}\ }\textbf {\bibinfo {volume}
  {5}},\ \bibinfo {pages} {03LT01} (\bibinfo {year} {2020})}\BibitemShut
  {NoStop}%
\bibitem [{Note1()}]{Note1}%
  \BibitemOpen
  \bibinfo {note} {We assume inversion and
  multiplication of generic $M\times M$ matrices has
  complexity $\protect \mathcal {O}(M^3)$. Optimizations like Strassens'
  algorithm reduce this complexity but do not affect the
  \protect \textit {relative} speedups presented in this
  work.}\BibitemShut {Stop}%
\bibitem [{\citenamefont {Nation}\ \emph {et~al.}(2021)\citenamefont {Nation},
  \citenamefont {Kang}, \citenamefont {Sundaresan},\ and\ \citenamefont
  {Gambetta}}]{PRXQuantum.2.040326}%
  \BibitemOpen
  \bibfield  {author} {\bibinfo {author} {\bibfnamefont {P.~D.}\ \bibnamefont
  {Nation}}, \bibinfo {author} {\bibfnamefont {H.}~\bibnamefont {Kang}},
  \bibinfo {author} {\bibfnamefont {N.}~\bibnamefont {Sundaresan}}, \ and\
  \bibinfo {author} {\bibfnamefont {J.~M.}\ \bibnamefont {Gambetta}},\ }\href
  {\doibase 10.1103/PRXQuantum.2.040326} {\bibfield  {journal} {\bibinfo
  {journal} {PRX Quantum}\ }\textbf {\bibinfo {volume} {2}},\ \bibinfo {pages}
  {040326} (\bibinfo {year} {2021})}\BibitemShut {NoStop}%
\bibitem [{\citenamefont {Yang}\ \emph {et~al.}(2022)\citenamefont {Yang},
  \citenamefont {Raymond},\ and\ \citenamefont {Uno}}]{PhysRevA.106.012423}%
  \BibitemOpen
  \bibfield  {author} {\bibinfo {author} {\bibfnamefont {B.}~\bibnamefont
  {Yang}}, \bibinfo {author} {\bibfnamefont {R.}~\bibnamefont {Raymond}}, \
  and\ \bibinfo {author} {\bibfnamefont {S.}~\bibnamefont {Uno}},\ }\href
  {\doibase 10.1103/PhysRevA.106.012423} {\bibfield  {journal} {\bibinfo
  {journal} {Phys. Rev. A}\ }\textbf {\bibinfo {volume} {106}},\ \bibinfo
  {pages} {012423} (\bibinfo {year} {2022})}\BibitemShut {NoStop}%
\bibitem [{\citenamefont {Schuld}\ and\ \citenamefont
  {Killoran}(2019)}]{Schuld2019a}%
  \BibitemOpen
  \bibfield  {author} {\bibinfo {author} {\bibfnamefont {M.}~\bibnamefont
  {Schuld}}\ and\ \bibinfo {author} {\bibfnamefont {N.}~\bibnamefont
  {Killoran}},\ }\href {\doibase 10.1103/PhysRevLett.122.040504} {\bibfield
  {journal} {\bibinfo  {journal} {Phys. Rev. Lett.}\ }\textbf {\bibinfo
  {volume} {122}},\ \bibinfo {pages} {040504} (\bibinfo {year}
  {2019})}\BibitemShut {NoStop}%
\bibitem [{\citenamefont {Huo}\ and\ \citenamefont {Li}(2022)}]{huo_2022}%
  \BibitemOpen
  \bibfield  {author} {\bibinfo {author} {\bibfnamefont {M.}~\bibnamefont
  {Huo}}\ and\ \bibinfo {author} {\bibfnamefont {Y.}~\bibnamefont {Li}},\
  }\href {\doibase 10.1103/PhysRevA.105.022427} {\bibfield  {journal} {\bibinfo
   {journal} {Phys. Rev. A}\ }\textbf {\bibinfo {volume} {105}},\ \bibinfo
  {pages} {022427} (\bibinfo {year} {2022})}\BibitemShut {NoStop}%
\bibitem [{\citenamefont {Peters}\ \emph {et~al.}(2022)\citenamefont {Peters},
  \citenamefont {Shyamsundar}, \citenamefont {Li},\ and\ \citenamefont
  {Perdue}}]{peters_2022}%
  \BibitemOpen
  \bibfield  {author} {\bibinfo {author} {\bibfnamefont {E.}~\bibnamefont
  {Peters}}, \bibinfo {author} {\bibfnamefont {P.}~\bibnamefont {Shyamsundar}},
  \bibinfo {author} {\bibfnamefont {A.~C.}\ \bibnamefont {Li}}, \ and\ \bibinfo
  {author} {\bibfnamefont {G.}~\bibnamefont {Perdue}},\ }\href {\doibase
  10.1103/PRXQuantum.3.040333} {\bibfield  {journal} {\bibinfo  {journal} {PRX
  Quantum}\ }\textbf {\bibinfo {volume} {3}},\ \bibinfo {pages} {040333}
  (\bibinfo {year} {2022})}\BibitemShut {NoStop}%
\bibitem [{\citenamefont {Mitarai}\ \emph {et~al.}(2018)\citenamefont
  {Mitarai}, \citenamefont {Negoro}, \citenamefont {Kitagawa},\ and\
  \citenamefont {Fujii}}]{Mitarai2018}%
  \BibitemOpen
  \bibfield  {author} {\bibinfo {author} {\bibfnamefont {K.}~\bibnamefont
  {Mitarai}}, \bibinfo {author} {\bibfnamefont {M.}~\bibnamefont {Negoro}},
  \bibinfo {author} {\bibfnamefont {M.}~\bibnamefont {Kitagawa}}, \ and\
  \bibinfo {author} {\bibfnamefont {K.}~\bibnamefont {Fujii}},\ }\href
  {\doibase 10.1103/PhysRevA.98.032309} {\bibfield  {journal} {\bibinfo
  {journal} {Phys. Rev. A}\ }\textbf {\bibinfo {volume} {98}},\ \bibinfo
  {pages} {032309} (\bibinfo {year} {2018})}\BibitemShut {NoStop}%
\bibitem [{\citenamefont {Khatri}\ \emph {et~al.}(2019)\citenamefont {Khatri},
  \citenamefont {LaRose}, \citenamefont {Poremba}, \citenamefont {Cincio},
  \citenamefont {Sornborger},\ and\ \citenamefont {Coles}}]{Khatri_2019}%
  \BibitemOpen
  \bibfield  {author} {\bibinfo {author} {\bibfnamefont {S.}~\bibnamefont
  {Khatri}}, \bibinfo {author} {\bibfnamefont {R.}~\bibnamefont {LaRose}},
  \bibinfo {author} {\bibfnamefont {A.}~\bibnamefont {Poremba}}, \bibinfo
  {author} {\bibfnamefont {L.}~\bibnamefont {Cincio}}, \bibinfo {author}
  {\bibfnamefont {A.~T.}\ \bibnamefont {Sornborger}}, \ and\ \bibinfo {author}
  {\bibfnamefont {P.~J.}\ \bibnamefont {Coles}},\ }\href {\doibase
  10.22331/q-2019-05-13-140} {\bibfield  {journal} {\bibinfo  {journal}
  {Quantum}\ }\textbf {\bibinfo {volume} {3}},\ \bibinfo {pages} {140}
  (\bibinfo {year} {2019})}\BibitemShut {NoStop}%
\bibitem [{\citenamefont {Geller}\ and\ \citenamefont
  {Sun}(2021)}]{Geller_2021}%
  \BibitemOpen
  \bibfield  {author} {\bibinfo {author} {\bibfnamefont {M.~R.}\ \bibnamefont
  {Geller}}\ and\ \bibinfo {author} {\bibfnamefont {M.}~\bibnamefont {Sun}},\
  }\href {\doibase 10.1088/2058-9565/abd5c9} {\bibfield  {journal} {\bibinfo
  {journal} {Quantum Sci. Technol.}\ }\textbf {\bibinfo {volume} {6}},\
  \bibinfo {pages} {025009} (\bibinfo {year} {2021})}\BibitemShut {NoStop}%
\bibitem [{\citenamefont {Wang}\ \emph {et~al.}(2021)\citenamefont {Wang},
  \citenamefont {Chen},\ and\ \citenamefont {Wang}}]{wang2021measurement}%
  \BibitemOpen
  \bibfield  {author} {\bibinfo {author} {\bibfnamefont {K.}~\bibnamefont
  {Wang}}, \bibinfo {author} {\bibfnamefont {Y.-A.}\ \bibnamefont {Chen}}, \
  and\ \bibinfo {author} {\bibfnamefont {X.}~\bibnamefont {Wang}},\ }\href@noop
  {} {\enquote {\bibinfo {title} {Measurement error mitigation via truncated
  neumann series},}\ } (\bibinfo {year} {2021}),\ \Eprint
  {http://arxiv.org/abs/2103.13856} {arXiv:2103.13856 [quant-ph]} \BibitemShut
  {NoStop}%
\bibitem [{\citenamefont {Lipton}\ \emph {et~al.}(1979)\citenamefont {Lipton},
  \citenamefont {Rose},\ and\ \citenamefont {Tarjan}}]{lipton1979generalized}%
  \BibitemOpen
  \bibfield  {author} {\bibinfo {author} {\bibfnamefont {R.~J.}\ \bibnamefont
  {Lipton}}, \bibinfo {author} {\bibfnamefont {D.~J.}\ \bibnamefont {Rose}}, \
  and\ \bibinfo {author} {\bibfnamefont {R.~E.}\ \bibnamefont {Tarjan}},\
  }\href@noop {} {\bibfield  {journal} {\bibinfo  {journal} {SIAM journal on
  numerical analysis}\ }\textbf {\bibinfo {volume} {16}},\ \bibinfo {pages}
  {346} (\bibinfo {year} {1979})}\BibitemShut {NoStop}%
\bibitem [{\citenamefont {Kılıç}\ and\ \citenamefont
  {Stanica}(2013)}]{KILIC2013126}%
  \BibitemOpen
  \bibfield  {author} {\bibinfo {author} {\bibfnamefont {E.}~\bibnamefont
  {Kılıç}}\ and\ \bibinfo {author} {\bibfnamefont {P.}~\bibnamefont
  {Stanica}},\ }\href {\doibase https://doi.org/10.1016/j.cam.2012.07.018}
  {\bibfield  {journal} {\bibinfo  {journal} {Journal of Computational and
  Applied Mathematics}\ }\textbf {\bibinfo {volume} {237}},\ \bibinfo {pages}
  {126} (\bibinfo {year} {2013})}\BibitemShut {NoStop}%
\bibitem [{\citenamefont {Lutkepohl}(1997)}]{lutkepohl1997handbook}%
  \BibitemOpen
  \bibfield  {author} {\bibinfo {author} {\bibfnamefont {H.}~\bibnamefont
  {Lutkepohl}},\ }\href@noop {} {\emph {\bibinfo {title} {Handbook of
  matrices}}},\ Vol.~\bibinfo {volume} {2}\ (\bibinfo {year} {1997})\ p.\
  \bibinfo {pages} {167}\BibitemShut {NoStop}%
\bibitem [{\citenamefont {Ozdaglar}\ \emph {et~al.}(2019)\citenamefont
  {Ozdaglar}, \citenamefont {Shah},\ and\ \citenamefont
  {Yu}}]{Lee2014SolvingFA}%
  \BibitemOpen
  \bibfield  {author} {\bibinfo {author} {\bibfnamefont {A.}~\bibnamefont
  {Ozdaglar}}, \bibinfo {author} {\bibfnamefont {D.}~\bibnamefont {Shah}}, \
  and\ \bibinfo {author} {\bibfnamefont {C.~L.}\ \bibnamefont {Yu}},\
  }\href@noop {} {\enquote {\bibinfo {title} {Asynchronous approximation of a
  single component of the solution to a linear system},}\ } (\bibinfo {year}
  {2019}),\ \Eprint {http://arxiv.org/abs/1411.2647} {arXiv:1411.2647 [cs.DS]}
  \BibitemShut {NoStop}%
\end{thebibliography}
%

\end{document}